\documentclass[journal]{IEEEtran}
\renewcommand{\figurename}{Fig.}
\usepackage{cite}

\ifCLASSINFOpdf
\else
\fi
\ifCLASSOPTIONcompsoc
  \usepackage[caption=false,font=normalsize,labelfont=sf,textfont=sf]{subfig}
\else
  \usepackage[caption=false,font=footnotesize]{subfig}
\fi
\usepackage[T1]{fontenc}
\usepackage[utf8]{luainputenc}

\usepackage{float}
\usepackage{graphicx}
\usepackage{epstopdf}
\usepackage{epsfig}
\AppendGraphicsExtensions{.tif}

\usepackage{gensymb}
\usepackage{algorithm}
\usepackage{algorithmicx}
\usepackage{amsmath}
\usepackage{amssymb}

\usepackage{threeparttable}
\usepackage{array}
\usepackage{xcolor}
\usepackage{bm}
\usepackage{mathrsfs}

\makeatletter

\providecommand{\tabularnewline}{\\}
\makeatother

\floatstyle{ruled}
\newfloat{algorithm}{tbp}{loa}
\providecommand{\algorithmname}{Algorithm}
\floatname{algorithm}{\protect\algorithmname}

\begin{document}
%
\title{Evaluating and Improving the Depth Accuracy of Kinect for Windows v2}
%
%
%

\author{Lin Yang, Longyu Zhang, Haiwei Dong, Abdulhameed Alelaiwi and Abdulmotaleb El Saddik,~\IEEEmembership{Fellow,~IEEE}

\thanks{L. Yang, L. Zhang, H. Dong, and A. El Saddik are with Multimedia Computing Research Laboratory (MCRLab), School of Electrical Engineering and Computer Science, University of Ottawa, 800 King Edward, Ottawa, ON, K1N 6N5, Canada (e-mail: hdong@uottawa.ca).}
\thanks{A. Alelaiwi is with Software Engineering Department, College of Computer and Information Sciences, King Saud University, Riyadh, 12372, Saudi Arabia.}
\thanks{Copyright (c) 2013 IEEE. Personal use of this material is permitted. However, permission to use this material for any other purposes must be obtained from the IEEE by sending a request to pubs-permissions@ieee.org.}
}

%
%

\markboth{IEEE Sensor Journal}%
{Shell \MakeLowercase{\textit{et al.}}: Bare Demo of IEEEtran.cls for Journals}
%


\maketitle

\begin{abstract}
Microsoft Kinect sensor has been widely used in many applications since
the launch of its first version. Recently, Microsoft released a new version of Kinect sensor with improved hardware. However, the accuracy assessment of the sensor remains to be answered. In this paper, we measure the depth accuracy of the newly released Kinect v2 depth sensor, and obtain a cone model to illustrate its accuracy distribution. We then evaluate the variance of the captured depth values by depth entropy. In addition, we propose a trilateration method to improve the depth accuracy with multiple Kinects simultaneously. The experimental results are provided to ascertain the proposed model and method.
\end{abstract}


%
\IEEEpeerreviewmaketitle

\section{Introduction}
%
%
%
%
\IEEEPARstart{I}{n} the past few years, the areas of computer vision have been developing
quickly due to the improvements in computer and camera capabilities.
While RGB cameras capture the color information, depth cameras measure the range information between the camera and the object, which
offers more convenience for three-dimensional (3D) model construction,
object tracking, movement detection, etc \cite{Nadia_2013}. Both
industrial and consumer-grade depth sensors have been developed to
satisfy the requirements from researchers and common users \cite{Dong_2014}. Typical
devices include Cyberware laser head scanner \cite{002Cyberware}, 
MESA Imaging SRFamily Time-of-Flight (ToF) camera \cite{TOFcam}, and PrimeSense Carmine structured-light cameras \cite{004Carmine}.

The launch of Microsoft Kinect for Windows sensor v1 \cite{001KinectWeb}
in November 2010 further enriched the depth-camera device options with its
low consumer price, compact size, and the capability to capture depth
and image data at video rate. Kinect v1 projects patterns consisting
of many stripes at once, and allows the acquisition of a multitude
of samples simultaneously. Kinect has several advantages: it can capture
color and depth images at video rate well in low light levels, resolve
silhouette ambiguities in pose, and is color and texture invariant.
Besides, its operation has no difference compared with video cameras,
making it be easily operated by common users \cite{008State}.

Recently, Microsoft has released Kinect v2, which much improved the depth measurement accuracy. Regarding the depth sensing principle, Kinect v1 adopts structured-light method, which projects patterns consisting of many stripes at once, or of arbitrary finges, and allow the acquisition of a multitude of samples simultaneously \cite{030Parts2}. Compared with v1, Kinect v2 utilized Time-of-Flight method, which use active sensors to measure the distance of a surface by calculating the round-trip time of a pulse of light \cite{129ToF}. As a result, the captured depth images from Kinect v2 have better quality, and in our work we mainly focus on evaluating the depth sensing capability of Kinect v2.

Besides evaluating Kinect v2's depth accuracy, we also propose a method to improve its accuracy by applying trilateration principle with multiple Kinects v2. The trilateration principle has been widely used in Global Positioning System (GPS), Wireless Sensor Networks (WSN), and other localization research areas \cite{023improved}. In our experiment, we put three Kinects at different locations while measuring the same target, and then applied the proposed trilateration method to achieve the optimal position of the target. The compelling results ascertained our method.

To the best of our knowledge, there have been no existing work evaluating the accuracy
of Kinect v2 until now. Thus, in this paper, we evaluate its accuracy and make
the following contributions:
\begin{itemize}
\item We measure the depth accuracy of the newly released Kinect v2 depth
sensor, and obtain a cone model to describe its accuracy
distribution.
\item We adopt entropy to evaluate the variance of the captured
depth values.
\item We propose a trilateration method to improve the depth accuracy with multiple Kinect v2
sensors.
\end{itemize}

The organization of this paper is as follows. Section II introduces
some researches related to depth camera accuracy evaluation. In
Section III, we explain the detailed method of the accuracy assessment.
Section IV illustrates our proposed trilateration method with multiple Kinects.
The experimental results and discussions are given in
Section V. Finally, we summarize the paper and draw the conclusion
in Section VI.

\section{Related Work}

A variety of approaches have been developed to evaluate the accuracy
of Kinect v1. Khoshelham et al. proposed
a mathematical model for Kinect v1 depth sensor measurement, and presented
a theoretical error analysis which provides an insight into the factors
affecting the accuracy of the data \cite{006Indoor,007Analysis}. In their model, they used calibration
parameters, such as focal length, principle point offsets, lens distortion
coefficients, base length, and the distance of the reference pattern,
to calculate the 3D coordinates. Their experimental results showed
that the random error of depth measurement becomes larger with the
increase in distance between the sensor and the target object. Another contribution of their work is that they adopted a point cloud obtained from a calibrated laser sensor as the ground truth to evaluate
the point cloud data reconstructed by Kinect v1 with careful registration.

As Microsoft Coorporation has developed a real-time
human skeletons recognition system by Kinect v1 \cite{011Parts}, some researchers also focused
on accuracy evaluation for the human body joints
tracking by Kinect. Wheat et al. asked the participants
to perform movements (such as reaching and throwing objects), and then compared
the detected joints movements by Kinect v1 and by the gold-standard
system (a 12 digital-camera capture system) \cite{009Establish}. From the results, they
concluded that Kinect v1 can be a potentially valuable motion analysis
tool, but certain improvements in accuracy are required. They also
tested the feasibility of Kinect v1 as a 3D scanning sensor and whole body
tracking device, and obtained good results. Galna et al. \cite{010Parkinson}
measured the relevant movements of subjects with Parkinson's disease
using Kinect v1. They found that Kinect v1 detects the timing of movement
repetitions accurately, but has varied success in measuring spatial
characteristics of movements based on the moving ranges.

Improving the accuracy of Kinect also attracts a lot of attentions. In the following part, we summarize previous work from two aspects, including calibrating color-depth sensors and increasing the overall performance with multiple Kinects:
\begin{itemize}

\item \textit{Calibration}: Kinect is able to capture both color and depth images. Thus, Herrera
et al. and Raposo et al. separately
proposed algorithms to calibrate the color-depth camera pair \cite{012Joint, 005Fast}.
Their disparity distortion correction model considerably improved
the reconstruction accuracy by taking into account color and depth
features simultaneously and considering the camera pair system as a whole.
With their contribution, the 3D models with better accuracy were reconstructed. 

\item \textit{Multiple Kinects}: As a single sensor may have limitations of resolution, field of view, or accuracy, some researchers used multiple sensors to improve their experimental results. For example, Tong et al. \cite{013Kinect3} proposed	a method using three Kinects simultaneously to scan 3D full human bodies. Their method overcame a single Kinect's comparably low X/Y resolution and depth accuracy drawbacks, which can result in low-quality acquired data. With careful global alignment, they solved the loop-closure problem efficiently.
\end{itemize}

In this paper, we proposed to apply trilateration principle with multiple Kinects v2 to improve the overall depth accuracy, which has not been conducted to the best of our knowledge. The trilateration principle has been widely used in many location-estimation-related research areas. For example, Awad et al. used Received Signal Strength Indicator (RSSI) in Wireless Sensor Networks (WSN) to realize adaptive distance estimation and location based on trilateration concept \cite{028awad2007adaptive}; Thomas et al. presented an alternative closed-form formulation to improve the robot localization result by trilateration \cite{029thomas2005revisiting}; and Bajaj et al. introduced several cost-effective Global Positioning Systems (GPS) which use three or more satellites to determine the target's latitude, longitude, and altitude with trilateration principle \cite{027bajaj2002gps}. In the following parts, we list the advantages of using trilateration principle in multi-Kinect's case:

\begin{figure}
\begin{centering}
\subfloat[\label{fig:Kinect-for-Windows-bef}]{\begin{centering}
\includegraphics[width=8cm]{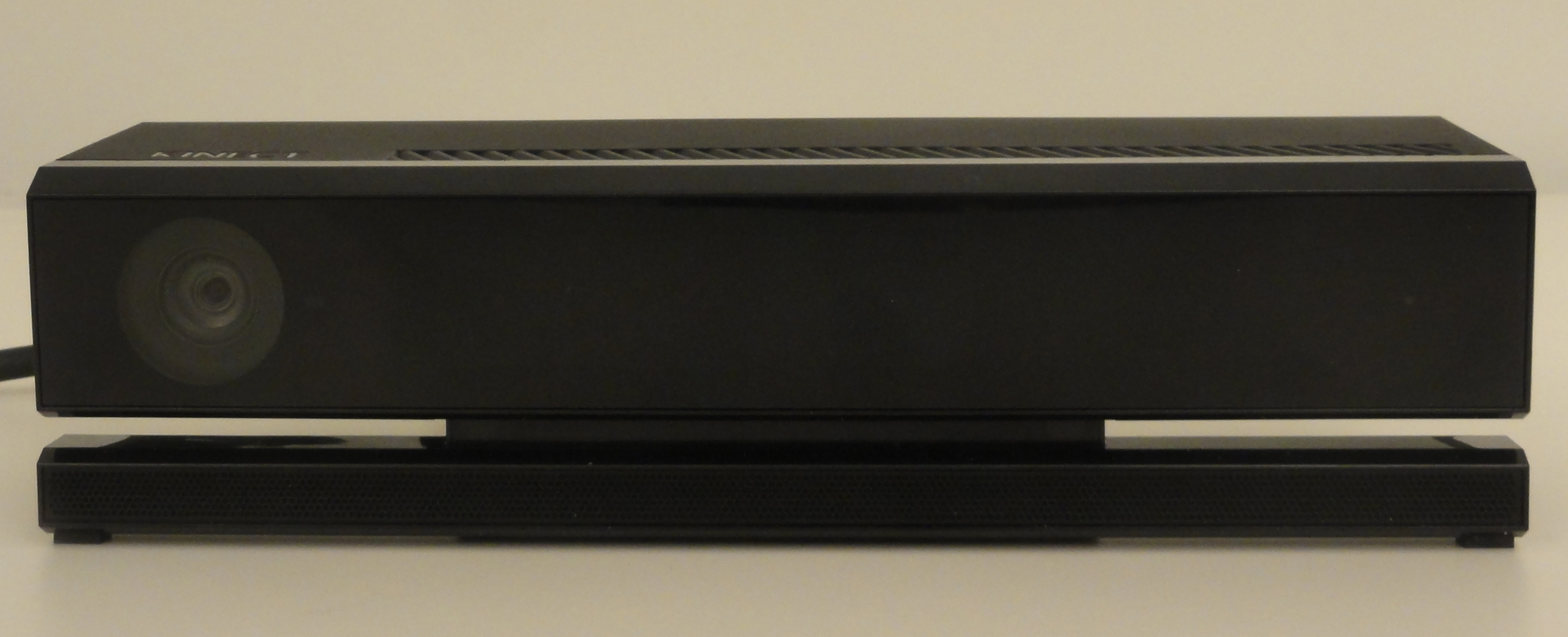}
\par\end{centering}
}
\par\end{centering}
\begin{centering}
\subfloat[\label{fig:Disassemble-2}]{\begin{centering}
\includegraphics[width=8cm]{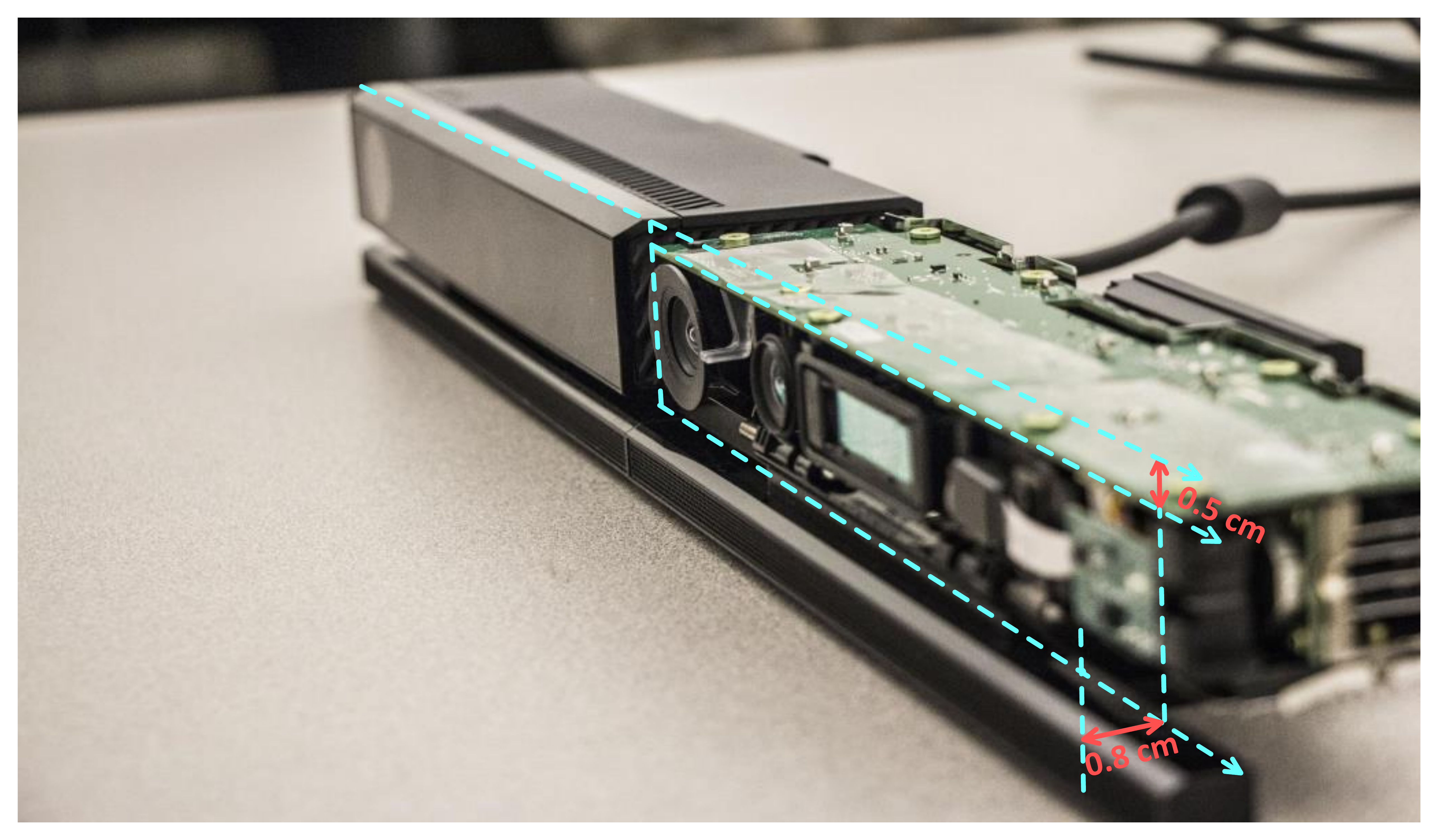}
\par\end{centering}
}
\par\end{centering}
\begin{centering}
\subfloat[\label{Disassemble Kinect}]{\begin{centering}
\includegraphics[width=8cm]{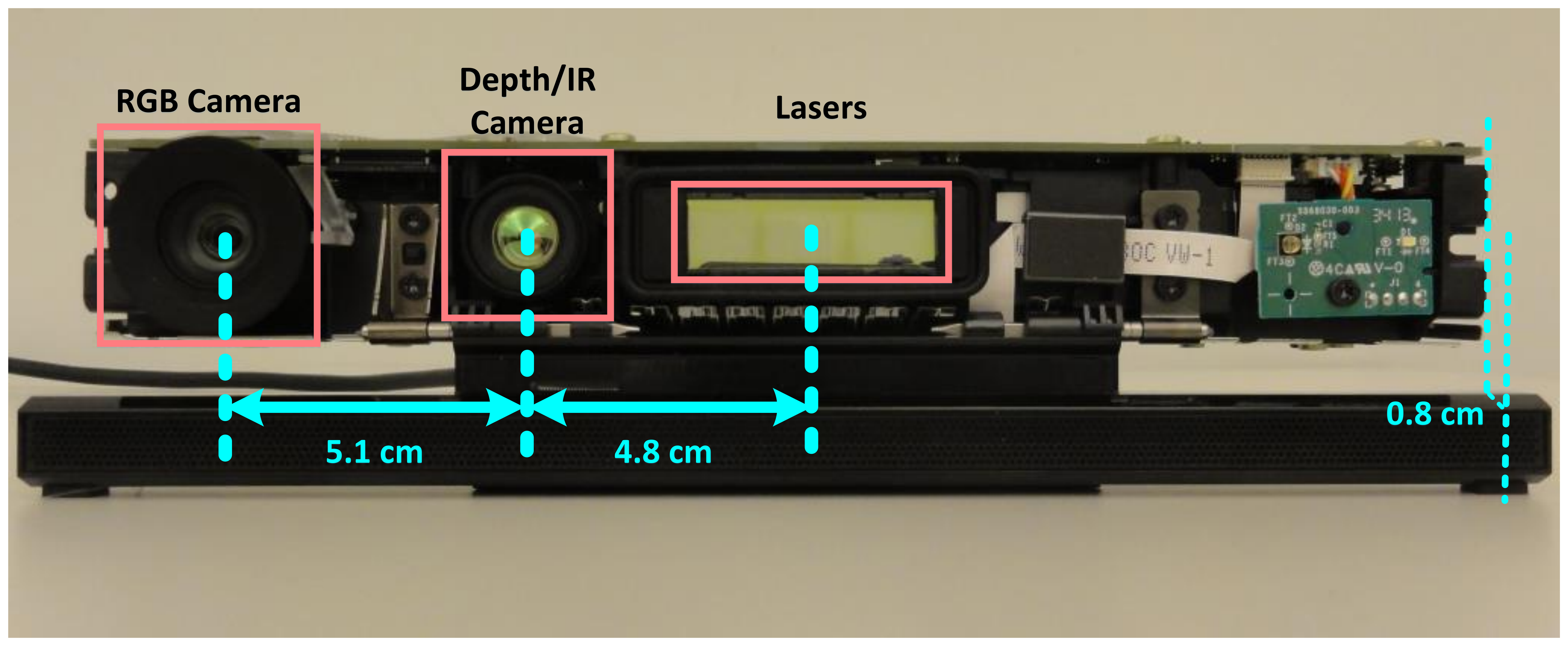}
\par\end{centering}
}
\par\end{centering}
\protect\caption{Microsoft Kinect for Windows v2 sensor. (a): Appearance of Kinect v2 (front view). (b): Camera configuration (isometric view). (c): Camera configuration (front view).}
\label{fig:Disassemble}
\end{figure}

First, trilateration principle considers each sensor and its corresponding measured distance as the center and radius of a sphere, and then calculates these spheres' common interaction point by solving the sphere equation set \cite{026efficient}. As the calculation is mainly based on the optimal estimation, the trilateration principle obtains much better position estimation compared with simply averaging the measurements from all of the sensors \cite{022relationship}.

Second, the solution framework of trilateration is quite flexible, and can be easily extended to the situation with numerous Kinects. Murphy et al. applied trilateration principle to track the trucks in a practical mining application with eight sensors \cite{024determination}.

Third, the redundancy views of multiple sensors can enhance the robustness and accuracy of the measurement results. Both experiments by Le et al. \cite{020le2009joint} and Beyer et al. \cite{021beyer2004practical} validated that additional sensors can reduce measurement errors.

\section{\label{Kinect Accuracy Assessment}Kinect Accuracy Assessment}
Since 3D data fusion, body tracking, and many other applications are based on the measured depth information, our accuracy assessment in this paper mainly focus on the depth camera embedded
in Kinect v2. Five attributes including accuracy distribution, depth resolution, depth entropy, edge noise and structural noise are evaluated to assess the performance of the depth camera. To localize the cameras' position of Kinect v2, we disassembled the sensor as shown in \figurename \ref{fig:Disassemble}.

To assess the performance of the depth camera embedded in Kinect v2, a 19-inch screen is used as a planar surface in the experiments of the following sections for collecting depth data. The screen is positioned perpendicular towards Kinect v2 or pointed with a specific angle towards Kinect v2. To guarantee the aforementioned specific angle (perpendicular case corresponds with 90 degree's case), a AGPtek Handheld Digital Laser Point Distance Meter (measuring range: 40m, accuracy: $\pm$2mm, laser class: class II, laser type: 635nm) was applied. Specifically, three distances between the Kinect front cover and the planar surface are taken with the mentioned laser distance meter. With these three distances and Kinect size data, the desired angle can be assured using geometrical relation.

\subsection{Accuracy Distribution}

\begin{figure*}
\begin{centering}
\subfloat[]{\begin{centering}
\includegraphics[width=5.8cm]{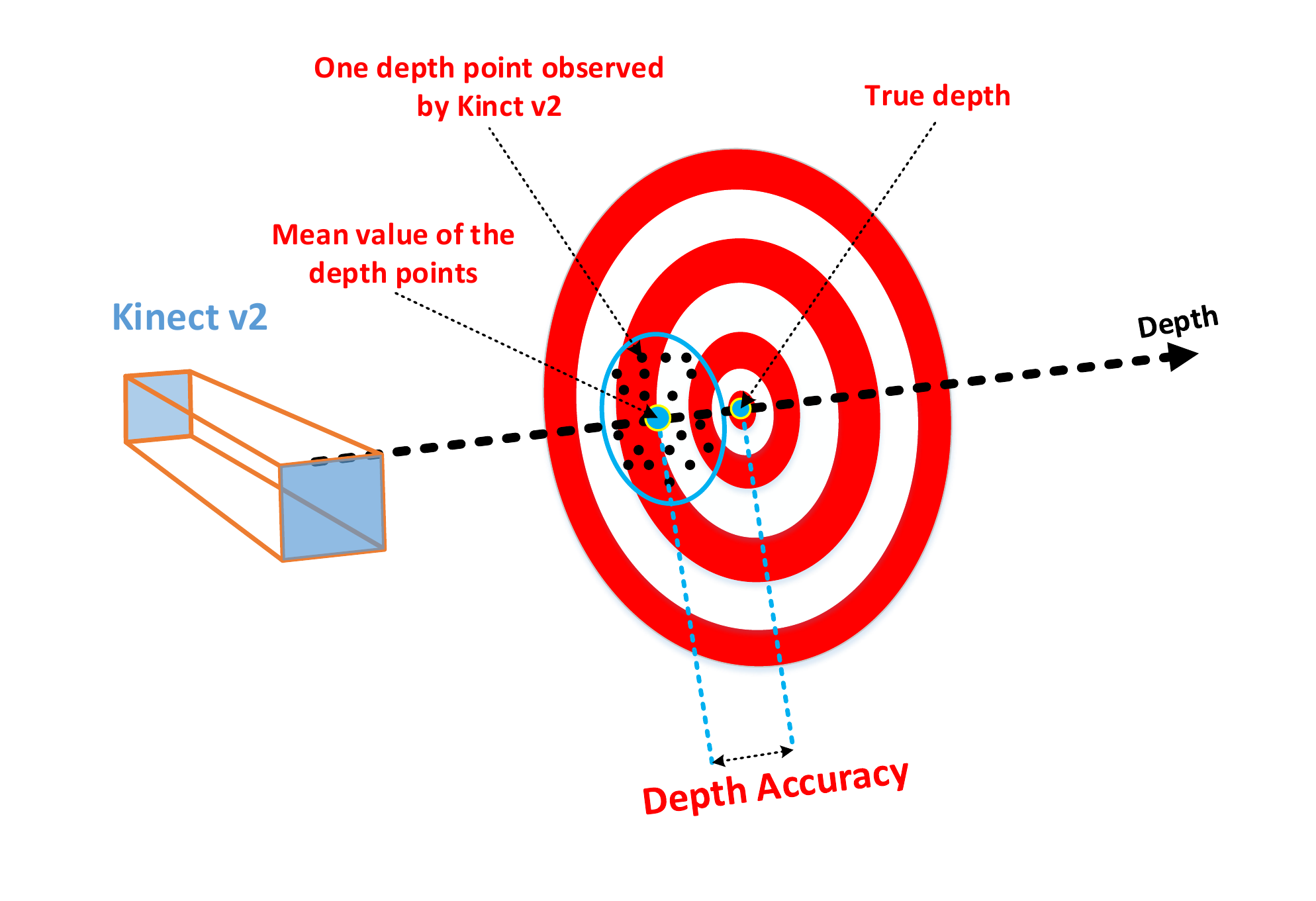}
\par\end{centering}

}\subfloat[]{\begin{centering}
\includegraphics[width=5.8cm]{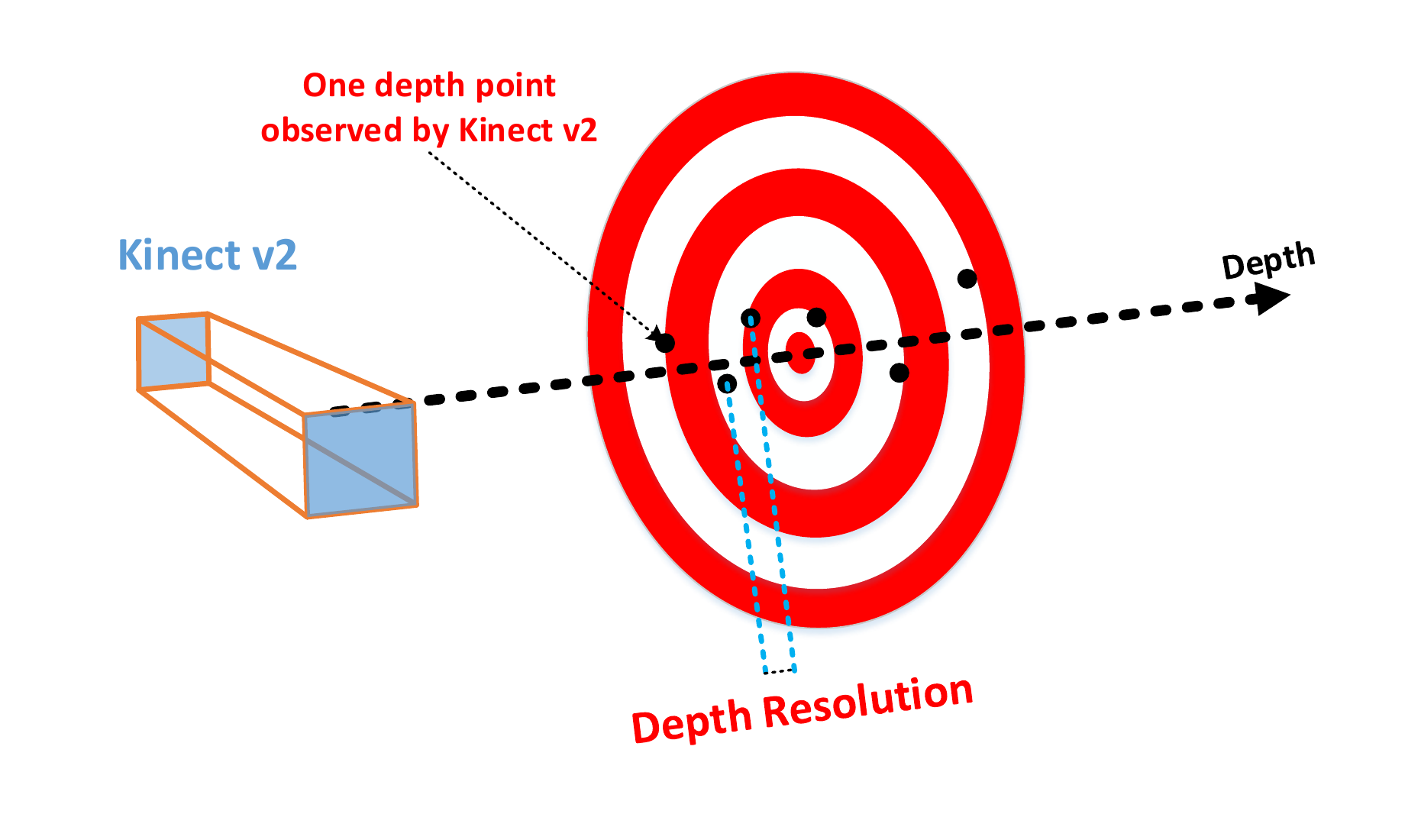}
\par\end{centering}

}\subfloat[]{\begin{centering}
\includegraphics[width=5.8cm]{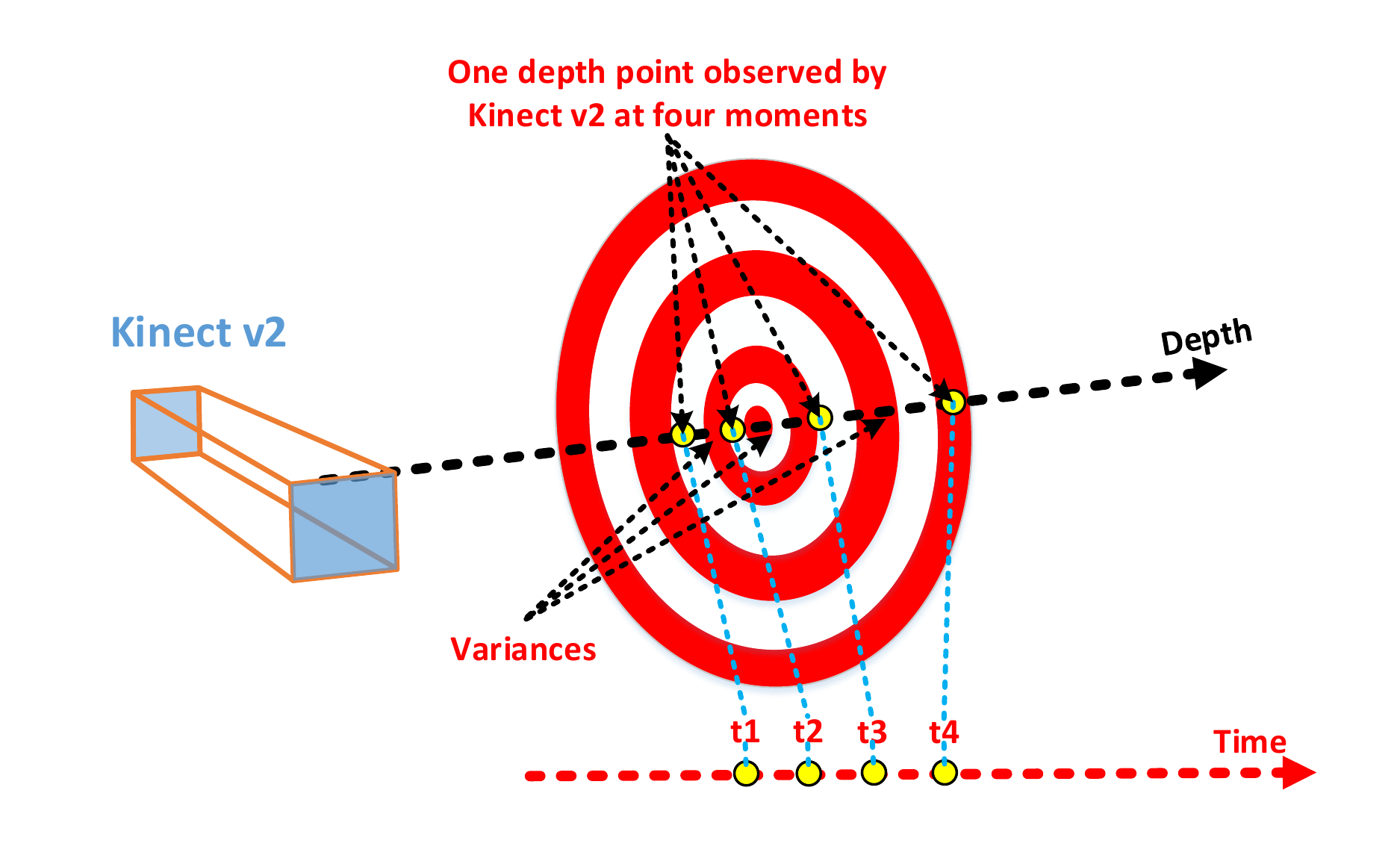}
\par\end{centering}

}
\par\end{centering}

\begin{centering}
\subfloat[]{\begin{centering}
\includegraphics[width=5.8cm]{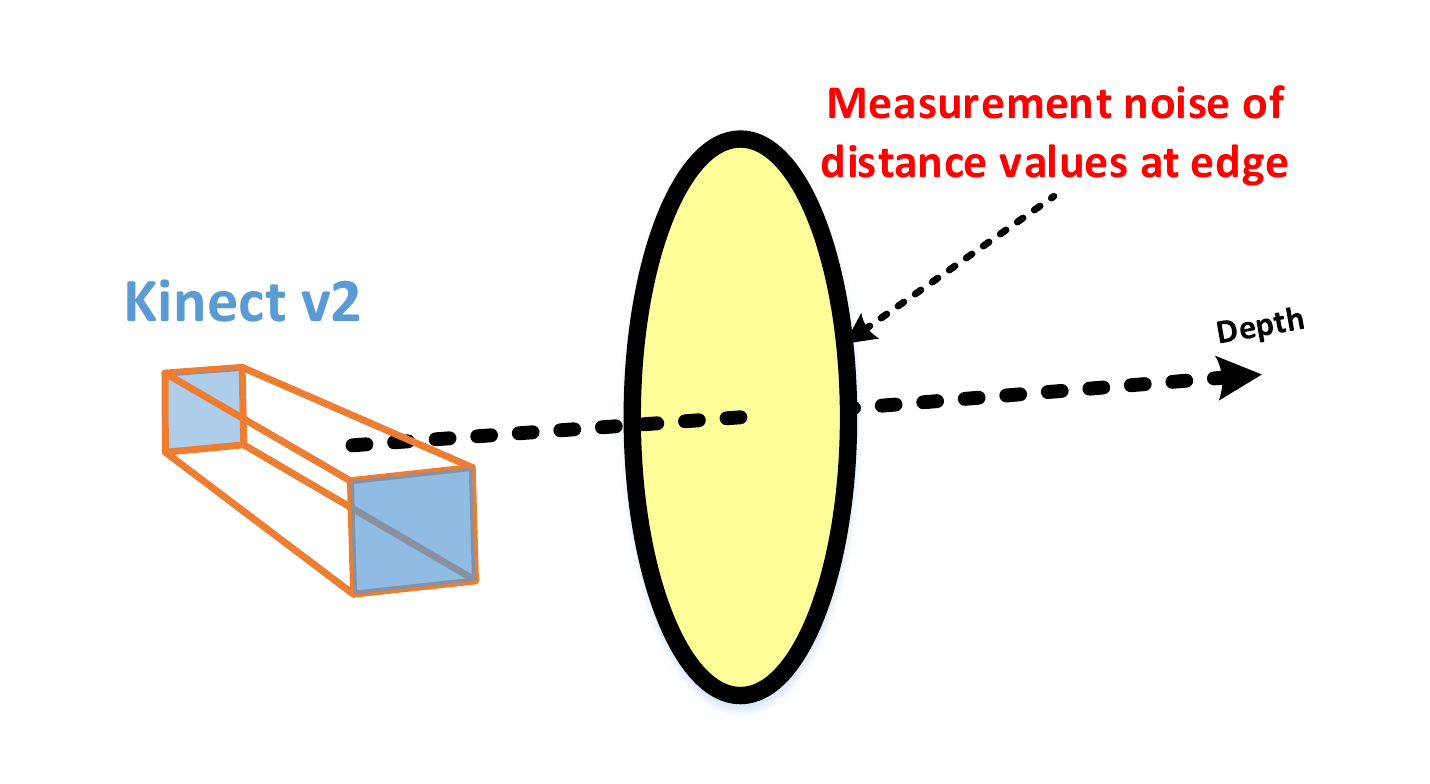}
\par\end{centering}

}\subfloat[]{\begin{centering}
\includegraphics[width=5.8cm]{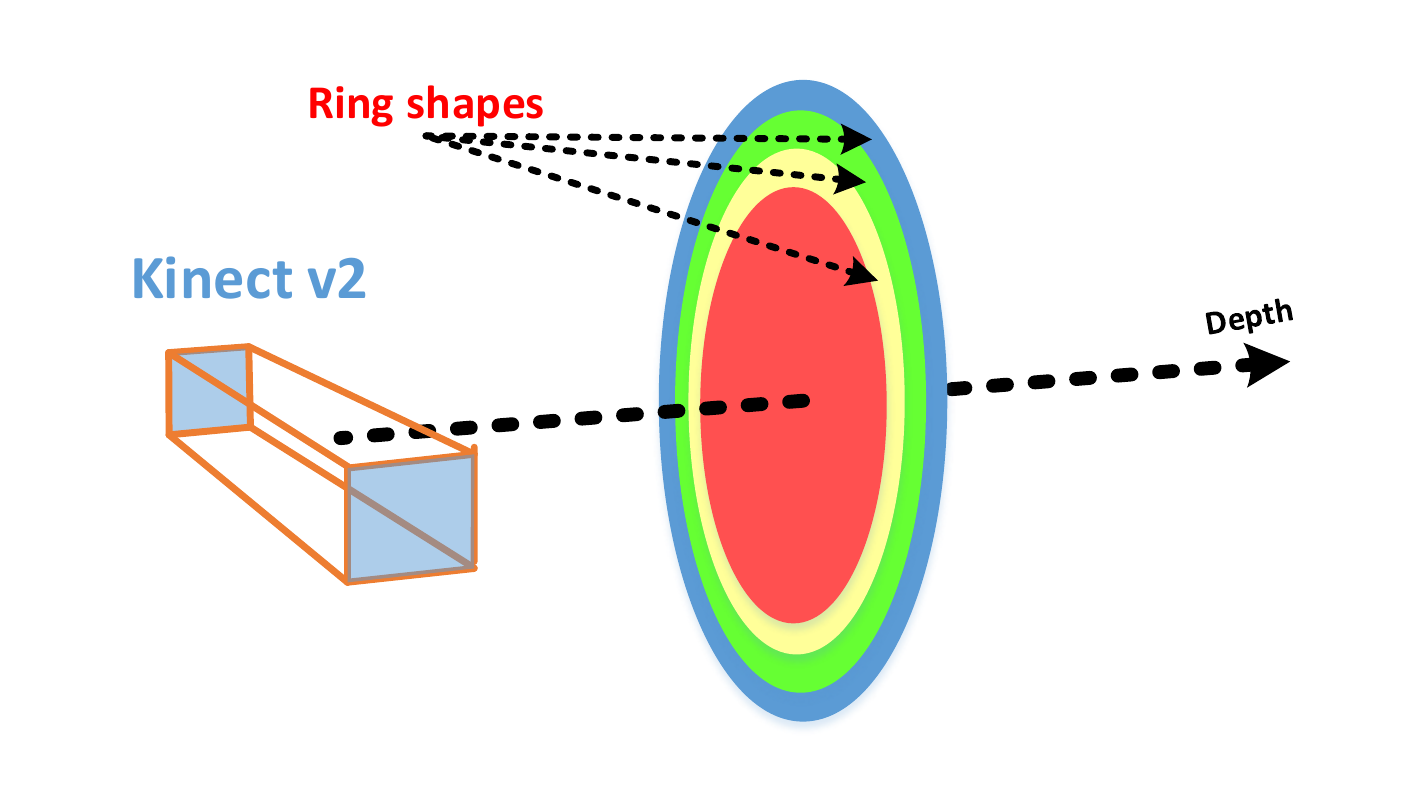}
\par\end{centering}

}
\par\end{centering}

\protect\caption{\label{fig:Concepts}Illustration of accuracy assessment of Kinect v2. (a) Depth accuracy. (b) Depth resolution. (c) Depth entropy. (d) Edge noise. (e) Structural noise. The target plates in (a-c) and (d-e) are parallel and perpendicular with the depth axis, respectively.}
\end{figure*}

In this paper, depth accuracy means the difference between the true depth value and the average value of the measured depth  values corresponding with a planar surface located in front of a Kinect v2. As shown in Fig.\ref{fig:Concepts}(a), the depth values inside the light blue circle are captured by a Kinect v2 representing a planar surface and a average depth value is calculated as the center of the circle. The difference between the mentioned average depth value and the true depth value is defined as depth accuracy.

Specifically, by denoting the depth values measured by a Kinect v2 as a set $M_{d}$, and the true distance between the planar surface and Kinect as $d$, the depth accuracy can be computed by the following equation

\begin{center}
\begin{equation} \label{eq:accuracy} \text{Depth Accuracy}=d- \text{mean}({M_{d}}) \end{equation}
\par\end{center}
where $Mean(\cdotp)$ computes the mean value of $(\cdotp)$. In the experiment, to compute the depth accuracy, a planar surface is placed perpendicular towards Kinect v2 with a specific distance, and the depth value matrix representing the planar surface is obtained by Kinect SDK. Meanwhile, the distance between Kinect v2 and the planar surface is measured by a laser distance meter, which is assumed as true value here.

To illustrate how the depth accuracy differs in the space, a 3D cone is built to show the accuracy distribution. Here the accuracy distribution are evaluated from both horizontal and vertical direction. In the horizontal direction, the planar surface is pointed perpendicular towards the Kinect v2. The planar surface is placed at the distance of 1m, 2m, 3m and 4m away from the Kinect sensor. In the vertical direction, we use a flat wall as the planar surface to evaluate the vertical accuracy and the Kinect sensor is positioned at the distance of 1m, 2m, 3m and 4m away from the wall. For both directions, at each location, a matrix of distance values in millimeters is exported by Kinect SDK (a Software Development Kit) and then loaded into Matlab for computation and visualization.

\subsection{Depth Resolution}

Depth resolution means the minimum detectable difference by the Kinect sensor in a certain continuous distance range. Thus, the smaller distance difference being detectable, the more precise depth values being measured, and  the higher resolution being achieved. As shown in Fig.\ref{fig:Concepts}(b), there are totally five depth values measured and the minimum difference among the differences between the adjacent depth values is the depth resolution. Specifically, the depth resolution is defined as follows
\begin{center}
\begin{equation} \label{eq:resolution} \text{Depth Resolution}= \min\left|M_d[i+1]-M_d[i]\right| \end{equation}
\par\end{center}
where $M_d[i+1]$ and $M_d[i]$ are two adjacent depth values in $M_d$. $\min(\cdotp)$ computes the minimum value of $(\cdotp)$. In our set-up, to compute depth resolution, a planar surface is positioned with two specific angle (45 or 60 degrees) towards Kinect v2. The experiment is repeated when the planar surface is positioned at different distances (1m, 1.5m, 2m, 2.5m, 3m, 3.5m and 4m away from the Kinect sensor). At each position, depth values, in the recorded frame, are extracted and processed accordingly.

\subsection{\label{Depth Entropy}Depth Entropy}

As limited to the performance of the depth camera, the infrared emitters, the surfaces'
materials and the ambient light, the depth values keep changing
over time. Even if the sensor's set-up is stationary, a given pixel may
vary in several millimeters. As shown in Fig.\ref{fig:Concepts}(c), the depth value at a specific position varies for times. By applying entropy concept, depth entropy is defined to illustrate the stability and reliability of the Kinect sensor with time. Specifically, the depth entropy is defined as follows
\begin{center}
\begin{equation} \label{eq:eq1} \text{Depth Entropy}(i)=-\sum_{j}p(i)_{j}\log_{2}p(i)_{j} \end{equation}
\par\end{center}
where $i$ is the index of the depth value in $M_d$. $j$ is the possible difference values of $M_d(i)$ between two adjacent time frame. $p(i)_j$ is frequency of occurrence of $j$ within the tested time frames. In our experiment set-up, a planar surface is pointed perpendicular towards the Kinect v2 with 1m, 2m, 3m, 4m distance away. The depth values representing the planar surface is exported from Kinect SDK each second and a total of 30 time frames are exported and post-processed to calculate the depth entropy.

\subsection{Edge Noise}
There are measurement noise occurring at the edges of an object when the object has a sharp contour. As shown in Fig.\ref{fig:Concepts}(d), a circular planar surface is positioned perpendicular towards the Kinect v2 where the colors represent different depth values measured by the Kinect v2. It is shown that the depth values corresponding with the contour have different values (i.e., measurement noise) compared with that corresponding with the central part. These measurement noise happened on the sharp contours of an object are defined as edge noise here. In our experiment set-up, a Kinect is pointed perpendicular towards a planar surface where a depth frame is recorded. The experiment is repeated with different locations of the planar surface (1m, 2m, 3m, and 4m away from the Kinect sensor, respectively). According to the depth values obtained from the frames, the edge noise's distribution is visualized and further analyzed on possible reasons.

\subsection{Structural Noise}
Ideally, the depth values should be constant
when a planar surface is pointed perpendicular towards the depth camera,
but in reality the camera cannot capture a perfect flat surface.  The depth values distribute in ring shapes with different radii as Fig.\ref{fig:Concepts}(e). This noise phenomenon is defined as structural noise. To assess this noise character, we assume that a wall is even enough and consider it as a flat surface. A Kinect v2 is pointed perpendicular towards the wall for capturing depth data representing the wall. This experiment is repeated with different Kinect v2 locations (1m, 2m 3m and 4m away from the Kinect sensor to the planar surface) to analyze this noise phenomenon.

\section{Multi-Kinect Trilateration}

\subsection{Trilateration Principle}
The basic idea of trilateration method is utilizing multiple sensors to measure the distances between the target and each sensor, and building several spheres by taking each sensor as a center point and its corresponding distance as the radius. Thus, the target position is the common interaction point of these spheres, which can be determined by solving a set of equations representing all the spheres \cite{024determination}. As the sensor number is usually much more than the to-be-determined position variables (i.e., the sensor information is usually redundant), the solution of the above mentioned equation set can be achieved under the least squares sense by applying pseudo-inverse. In this paper, we apply this trilateration principle to improve the overall accuracy of the multi-Kinect's measurements.

\subsection{\label{Trilateration for Improving Multi-Kinect Accuracy}Trilateration for Improving Multi-Kinect Accuracy}

\begin{figure*}
\begin{centering}
\includegraphics[width=18cm]{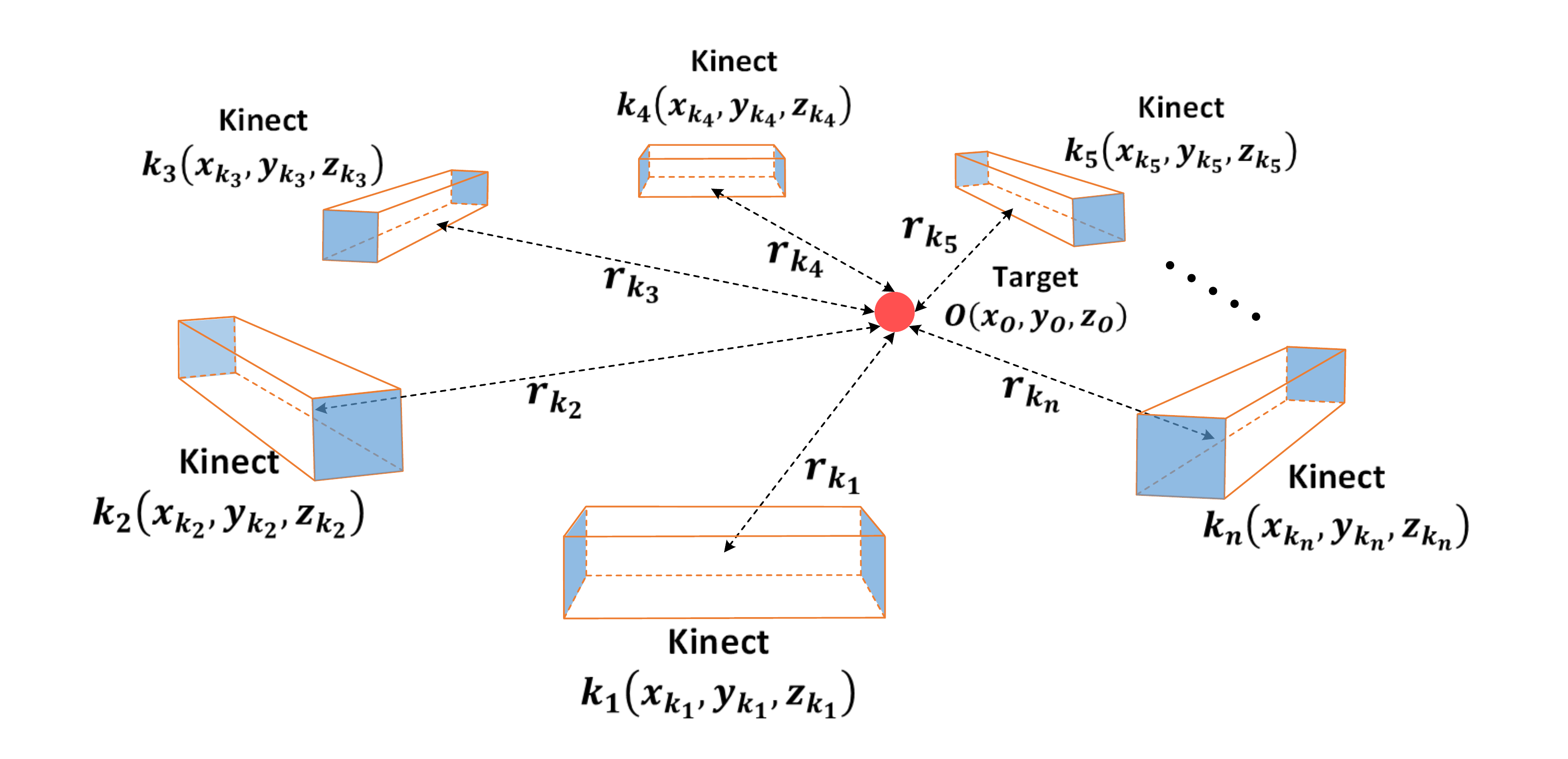}
\par\end{centering}
\begin{centering}
\protect\caption{\label{fig:Trilateration}Coordinate setting of the multi-Kinect trilateration.}
\par\end{centering}
\end{figure*}

Suppose there are $n$ Kinect v2 sensors positioned accordingly in a specific coordinate system and the target is placed where each Kinect sensor is able to observe it (Fig.\ref{fig:Trilateration}). The target position can be calculated by solving the following sphere equations

\begin{equation}
\label{eq:trilateration}
 \left\{
\begin{array}{c}
(x_{o}-x_{k_{1}})^{2}+(y_{o}-y_{k_{1}})^{2}+(z_{o}-z_{k_{1}})^{2}=r_{k_{1}}^{2}\\
\vdots\\
(x_{o}-x_{k_{i}})^{2}+(y_{o}-y_{k_{i}})^{2}+(z_{o}-z_{k_{i}})^{2}=r_{k_{i}}^{2}\\ 
\vdots\\
(x_{o}-x_{k_{n}})^{2}+(y_{o}-y_{k_{n}})^{2}+(z_{o}-z_{k_{n}})^{2}=r_{k_{n}}^{2}\\
\end{array}
\right.
\end{equation}
where $(x_{k_{i}},\:y_{k_{i}},\:z_{k_{i}})$ is the location of Kinect $k_{i}$. $(x_{o},\:y_{o},\:z_{o})$ is the target position to be solved. $r_{k_{i}}$ is the range between Kinect $k_{i}$ and the target position measured by Kinect $k_{i}$. However, there is always noises and disturbances in real applications (such as accuracy limit for hardware devices, the clock offsets, etc.). Here, we use $e_{k_{i}}^2$ to represent the noise coming from $i$-th Kinect and the Equation \eqref{eq:trilateration} can be written as

\begin{equation}
\label{eq:radiuses}
 \left\{
\begin{array}{c}
(x_{o}-x_{k_{1}})^{2}+(y_{o}-y_{k_{1}})^{2}+(z_{o}-z_{k_{1}})^{2}-r_{k_{1}}^{2}=e_{k_{1}}^2\\
\vdots\\
(x_{o}-x_{k_{i}})^{2}+(y_{o}-y_{k_{i}})^{2}+(z_{o}-z_{k_{i}})^{2}-r_{k_{i}}^{2}=e_{k_{i}}^2\\ 
\vdots\\
(x_{o}-x_{k_{n}})^{2}+(y_{o}-y_{k_{n}})^{2}+(z_{o}-z_{k_{n}})^{2}-r_{k_{n}}^{2}=e_{k_{n}}^2\\
\end{array}
\right.
\end{equation}

Hence, we can obtain an optimal position of the target by minimizing the sum of the errors $e_{k_{i}}^2$ ($1\leq i \leq n$)

\begin{center}
\begin{equation} \label{eq:minimumE} (x_{o},\: y_{o},\: z_{o})=\arg\min(\sum_{i=1}^{n}e_{k_{i}}^{2}) \end{equation}
\par\end{center}

From Equation \eqref{eq:trilateration}, for the $i$-th Kinect, the sphere equation containing the range measurement $r_{k_i}$ and its position coordinate $(x_{k_i},y_{k_i},z_{k_i})$ is
\begin{center}
\begin{equation}(x_{o}-x_{k_{i}})^{2}+(y_{o}-y_{k_{i}})^{2}+(z_{o}-z_{k_{i}})^{2}=r_{k_{i}}^{2} \end{equation}
\par\end{center}
where $i=1,2,\cdots,n$. By adding and subtracting $x_{k_{j}}$, $y_{k_{j}}$ and $z_{k_{j}}$ ($j=1,2,\cdots,n$, $j\neq i$) to the above equation, we have
\begin{center}
\begin{equation}
\begin{aligned}
&(x_{o}-x_{k_{j}}+x_{k_{j}}-x_{k_{i}})^{2}+
(y_{o}-y_{k_{j}}+y_{k_{j}}-y_{k_{i}})^{2}\\
+&(z_{o}-z_{k_{j}}+z_{k_{j}}-z_{k_{i}})^{2}=r_{k_{i}}^{2} 
\end{aligned}
\end{equation}
\par\end{center}
After expanding and merging terms, we obtain

\begin{center}
\begin{equation} 
\label{eq:trilateration2}
\begin{aligned}
&\quad\,(x_{o}-x_{k_{j}})(x_{k_{i}}-x_{k_{j}})+(y_{o}-y_{k_{j}})(y_{k_{i}}-y_{k_{j}})\\
&+(z_{o}-z_{k_{j}})(z_{k_{i}}-z_{k_{j}})\\
=&\frac{1}{2}[(x_{o}-x_{k_{j}})^{2}+(y_{o}-y_{k_{j}})^{2}+(z_{o}-z_{k_{j}})^{2}-r_{k_{i}}^{2}\\
&+(x_{k_{i}}-x_{k_{j}})^{2}+(y_{k_{i}}-y_{k_{j}})^{2}+(\text{z}_{k_{i}}-z_{k_{j}})^{2}]\\
=&\frac{1}{2}(r_{k_{j}}^{2}-r_{k_{i}}^{2}+l_{k_{i}k_{j}}^{2})=b_{ij}
\end{aligned}
\end{equation}
\par\end{center}
where $l_{k_{i}k_{j}}$ is the distance between Kinect $k_{i}$ and Kinect $k_{j}$, i.e.

\begin{center}
\begin{equation} 
l_{k_{i}k_{j}}=((x_{k_{i}}-x_{k_{j}})^{2}+(y_{k_{i}}-y_{k_{j}})^{2}+(\text{z}_{k_{i}}-z_{k_{j}})^{2})^\frac{1}{2}
\end{equation}
\par\end{center}
For Equation \eqref{eq:trilateration2}, let $i=2, \cdots, n$, $j=1$ and thus, Equation \eqref{eq:trilateration} is reformulated as $n-1$ linear equations with three unknown variables ($x_{o}$, $y_{o}$ and $z_{o}$)

\begin{equation} 
 \left\{
\begin{array}{lll}
(x_{o}-x_{k_{1}})(x_{k_{2}}-x_{k_{1}})+(y_{o}-y_{k_{1}})(y_{k_{2}}-y_{k_{1}})\\
+(z_{o}-z_{k_{1}})(z_{k_{2}}-z_{k_{1}})=\frac{1}{2}(r_{k_{1}}^{2}-r_{k_{2}}^{2}+l_{k_{2}k_{1}}^{2})=&b_{21}\\
&\\
(x_{o}-x_{k_{1}})(x_{k_{3}}-x_{k_{1}})+(y_{o}-y_{k_{1}})(y_{k_{3}}-y_{k_{1}})\\
+(z_{o}-z_{k_{1}})(z_{k_{3}}-z_{k_{1}})=\frac{1}{2}(r_{k_{1}}^{2}-r_{k_{3}}^{2}+l_{k_{3}k_{1}}^{2})=&b_{31}\\
\qquad \qquad \qquad \qquad \qquad \vdots  &\vdots \\
(x_{o}-x_{k_{1}})(x_{k_{n}}-x_{k_{1}})+(y_{o}-y_{k_{1}})(y_{k_{n}}-y_{k_{1}})\\
+(z_{o}-z_{k_{1}})(z_{k_{n}}-z_{k_{1}})=\frac{1}{2}(r_{k_{1}}^{2}-r_{k_{n}}^{2}+l_{k_{n}k_{1}}^{2})=&b_{n1}\\
\end{array}
\right.
\end{equation}
which can be written in the form of a general linear equation set
\begin{center}
\begin{equation} 
\label{eq:matrix1}
\mathbf{A}\mathbf{x}=\mathbf{b}
\end{equation}
\par\end{center}
where
\begin{equation}
\begin{aligned}
\mathbf{A}=&\left[\begin{array}{ccc}
x_{k_{2}}-x_{k_{1}} & y_{k_{2}}-y_{k_{1}} & z_{k_{2}}-z_{k_{1}}\\
x_{k_{3}}-x_{k_{1}} & y_{k_{3}}-y_{k_{1}} & z_{k_{3}}-z_{k_{1}}\\
\vdots & \vdots & \vdots\\
x_{k_{n}}-x_{k_{1}} & y_{k_{n}}-y_{k_{1}} & z_{k_{n}}-z_{k_{1}}
\end{array}\right],\\
\mathbf{x}=&\left[\begin{array}{c}
x_{o}-x_{k_{1}}\\
y_{o}-y_{k_{1}}\\
z_{o}-z_{k_{1}}
\end{array}\right],\;
\mathbf{b}=\left[\begin{array}{c}
b_{21}\\
b_{31}\\
\vdots\\
b_{n1}
\end{array}\right]
\end{aligned}
\end{equation}
Equation \eqref{eq:matrix1} can be solved by
\begin{center}
\begin{equation}
\mathbf{x}=\mathbf{A}^{+}\mathbf{b}=(\mathbf{A}^{T}\mathbf{A})^{-1}\mathbf{A}^{T}\mathbf{b}
\end{equation}
\par\end{center}
where $\bm{A}^{+}$ is the pseudo-inverse of matrix $\bm{A}$. According to matrix theory, the above solution provides the optimal target position under the sense of least squares. In other words, this solution provides the best estimation of the target position which minimizes the sum of the residual errors coming from the sphere equations \cite{MarixA_1990}. It is noted that, if the matrix $\bm{A}^{T}\bm{A}$ is singular or ill-conditioned, singular value decomposition (SVD)  can be applied to calculated the pseudo-inverse of $\bm{A}$.

\subsection{A Case Study}
\begin{figure*}
\begin{centering}
\subfloat[]{\begin{centering}
\includegraphics[width=10cm]{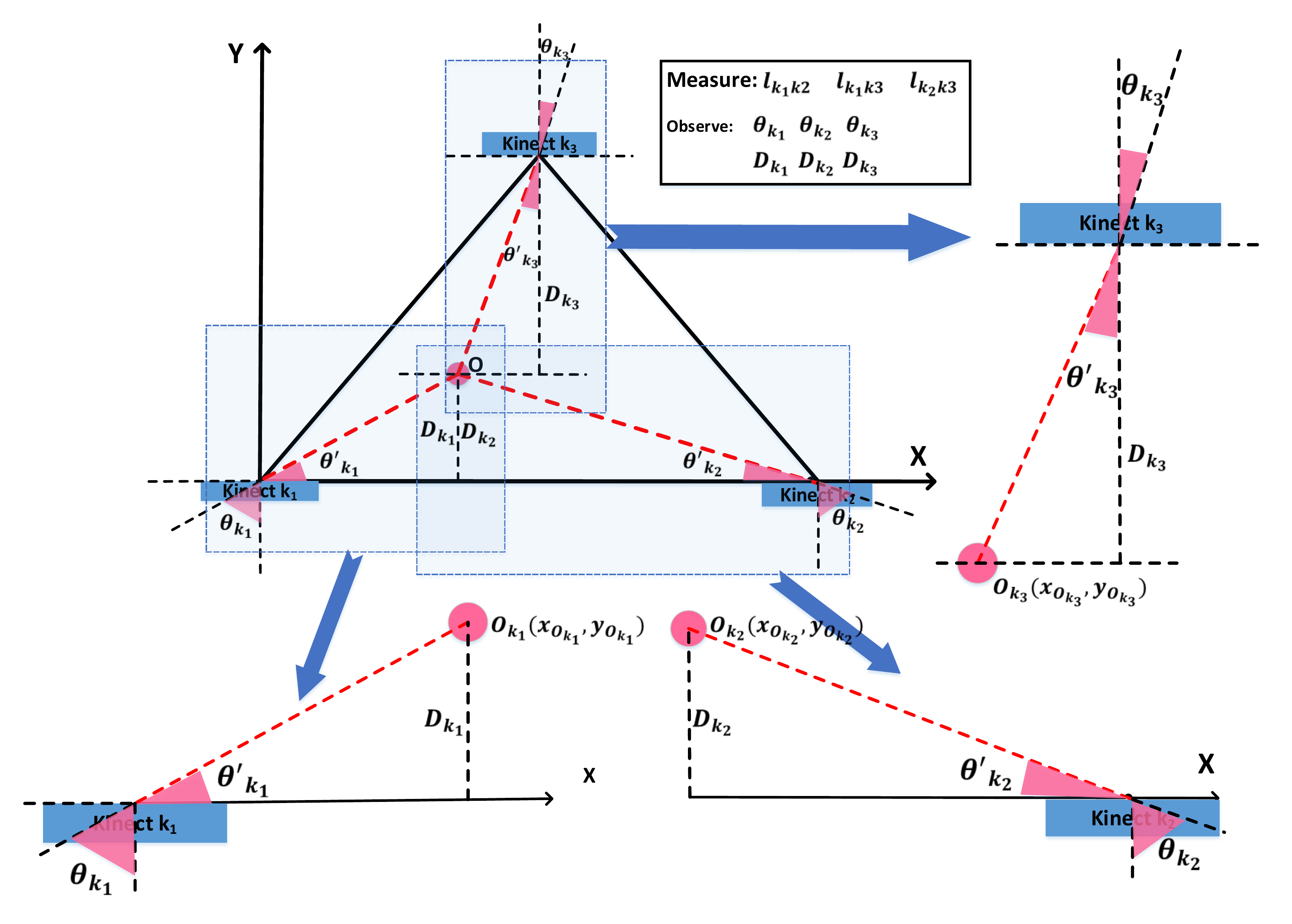}
\par\end{centering}

}\subfloat[]{\begin{centering}
\includegraphics[width=8cm]{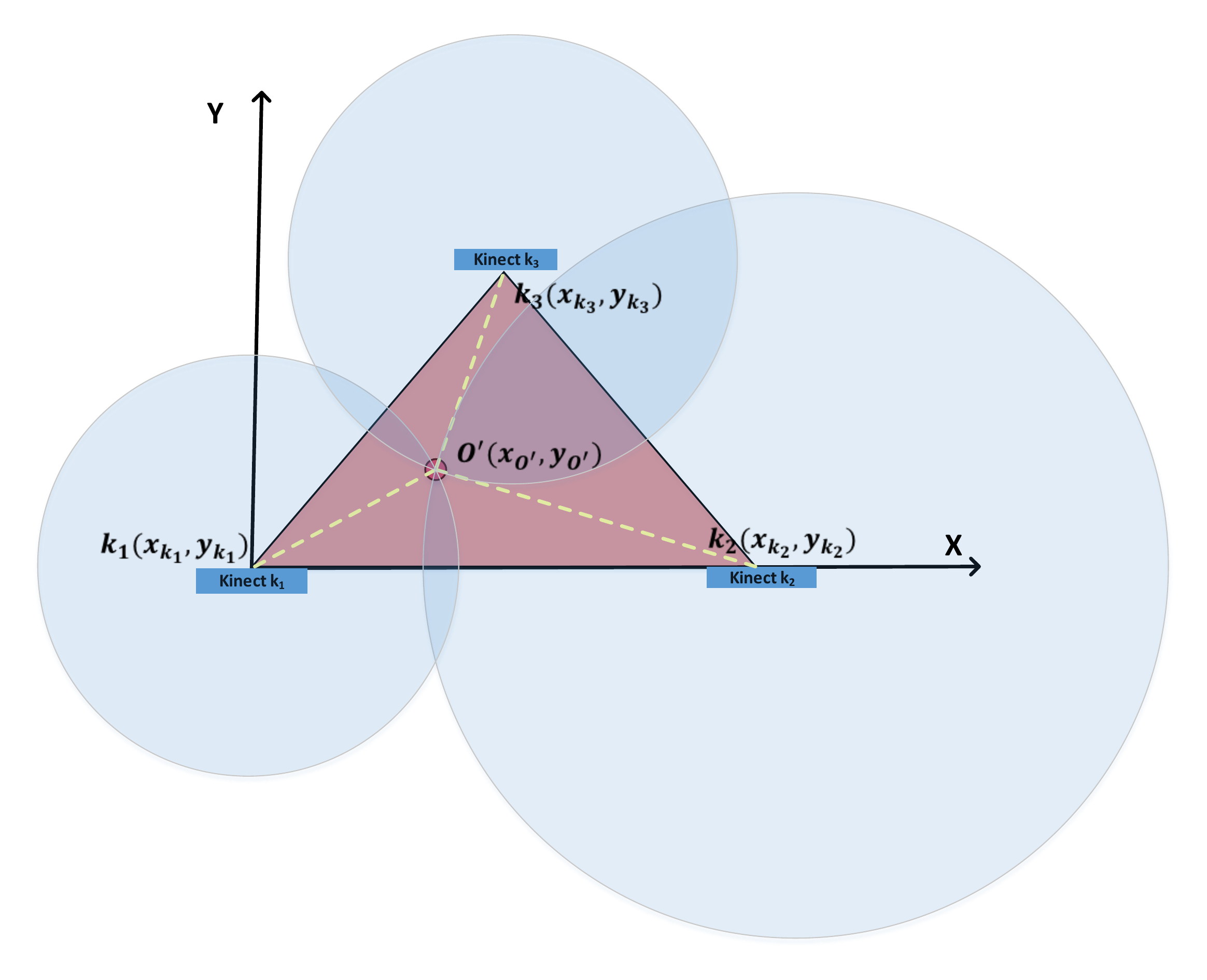}
\par\end{centering}

}
\par\end{centering}
\protect\caption{\label{fig:multi-Kinect}Multi-Kinect trilateration. (a) Trilateration set-up of multi-Kinect within a coordinate system. (b) The intersection point ($O^{'}$) of spheres centered with Kinect $k_{1}$, $k_{2}$ and $k_{3}$.}
\end{figure*}

We apply trilateration principle to three-Kinect set-up to improve the localization accuracy
of a target object when it is placed where the Kinect sensor cannot generate depth values accurately (e.g. distance larger than 4m from the observing sensor). In our configuration, the target object is a planar surface located more than 4m away from the three
Kinect sensors. The set-up is shown in \figurename\ref{fig:multi-Kinect}(a). In this figure, the three vertexes ($k_{1}$, $k_{2}$ and $k_{3}$) represent the locations of the three Kinects and $O$ represents the location of the target object.
$l_{k_{1}k_{2}}$, $l_{k_{1}k_{3}}$, $l_{k_{2}k_{3}}$ are measured in real world, where $l_{k_{i}k_{j}}=||k_{i}-k_{j}||_2$ is the distance between Kinect $k_{i}$ and Kinect $k_{j}$. The depth-angle pair $(\theta_{k_{1}}, \:D_{k_{1}})$, $(\theta_{k_{1}}, \:D_{k_{2}})$ and $(\theta_{k_{3}}, \:D_{k_{1}})$ are measured by Kinect $k_{1}$, $k_{2}$ and $k_{3}$, respectively. The depth values $D_{k_{1}}$,  $D_{k_{2}}$ and $D_{k_{3}}$ are calculated by averaging 20$\times$20 pixels' values
at the center of the planar surface in the depth image from each Kinect.

The trilateration method for this case study is divided into three steps where the input is the measured parameters of $l_{k_{1}k_{2}}$, $l_{k_{1}k_{3}}$, $l_{k_{2}k_{3}}$ and the observed parameters $(\theta_{k_{1}}, \:D_{k_{1}})$, $(\theta_{k_{2}}, \:D_{k_{2}})$, $(\theta_{k_{3}}, \:D_{k_{3}})$ from Kinect $k_{1}$, $k_{2}$ and $k_{3}$. The output is the optimized position of $O$ (denoted as $O^{'}$). According to geometry rules, the detailed procedures are summarized as follows:

\textit{Step 1}: We calculate the vertexes' positions $(k_{1}(x_{k_{1}},\, y_{k_{1}})$, $k_{1}(x_{k_{2}},\, y_{k_{2}})$, $k_{1}(x_{k_{3}},\, y_{k_{3}}))$ as the centers of three spheres with measured parameters ($l_{k_{1}k_{2}}$, $l_{k_{1}k_{3}}$, $l_{k_{2}k_{3}}$)
\begin{equation}
 \left\{
\begin{array}{l}
x_{k_{1}}=0\\
y_{k_{1}}=0\\
x_{k_{2}}=l_{k_{1}k_{2}}\\
y_{k_{2}}=0\\
x_{k_{3}}=\sqrt{l_{k_{1}k_{3}}^{2}-y_{k_{3}}^{2}}, \\
y_{k_{3}}=\frac{2\sqrt{S_{1}(S_{1}-l_{k_{1}k_{2}})(S_{1}-l_{k_{1}k_{3}})(S_{1}-l_{k_{2}k_{3}})}}{l_{k_{1}k_{2}}} \\
\end{array}
\right.
\end{equation}
where $S_{1}=\frac{1}{2}(l_{k_{1}k_{2}}+l_{k_{1}k_{3}}+l_{k_{1}k_{2}})$.

\textit{Step 2}: We calculate the angles and radiuses with parameters exported from each Kinect v2. The angles (from specific perspective of each Kinect v2) are calculated by Equation \eqref{eq:angle}
\begin{equation}
 \left\{
\label{eq:angle}
\begin{array}{l}
\theta_{k_{1}}^{'}=\frac{\pi}{2}-\theta_{k_{1}}\\
\theta_{k_{2}}^{'}=\frac{\pi}{2}-\theta_{k_{2}}\\
\theta_{k_{3}}^{'}=\theta_{k_{3}}\\
\end{array}
\right.
\end{equation}
In addition, we compute the three distances from the target object (location of $O$) to each Kinect sensor as three radiuses by Equation \eqref{eq:radiuses}
\begin{equation}
\label{eq:radiuses}
 \left\{
\begin{array}{l}
r_{k_{1}}=\frac{D_{k_{1}}}{sin\theta_{k_{1}}^{'}}\\
r_{k_{2}}=\frac{D_{k_{2}}}{sin\theta_{k_{1}}^{'}}\\ 
r_{k_{3}}=\frac{D_{k_{3}}}{sin\theta_{k_{1}}^{'}}\\
\end{array}
\right.
\end{equation}

\textit{Step 3}: The position of $O$ is computed with trilateration method which is denoted as $O^{'}(x_{O^{'}},\, y_{O^{'}})$ in \figurename\ref{fig:multi-Kinect}(b), by solving Equation \eqref{eq:spheres}
\begin{equation}
\label{eq:spheres}
 \left\{
\begin{array}{c}
(x_{O^{'}}-x_{k_{1}})^{2}+(y_{O^{'}}-y_{k_{1}})^{2}+(z^{'}-z_{k_{1}})^{2}=r_{k_{1}}^{2} \\
(x_{O^{'}}-x_{k_{2}})^{2}+(y_{O^{'}}-y_{k_{2}})^{2}+(z^{'}-z_{k_{2}})^{2}=r_{k_{2}}^{2} \\
(x_{O^{'}}-x_{k_{3}})^{2}+(y_{O^{'}}-y_{k_{3}})^{2}+(z^{'}-z_{k_{3}})^{2}=r_{k_{3}}^{2} \\
\end{array}
\right.
\end{equation}
where $z_{k_{1}}=z_{k_{2}}=z_{k_{3}}=0$, as Kinect v2 $k_{1}$, $k_{2}$ and $k_{3}$ are set at the same horizontal plane. As it is clarified in Section \ref{Trilateration for Improving Multi-Kinect Accuracy}, the three equations can be reformulated in matrix form (Equation \eqref{eq:matrix1}) where

\begin{equation}
\begin{aligned}
\mathbf{A}=&\begin{bmatrix}l_{k_{1}k_{2}} & 0\\
\sqrt{l_{k_{1}k_{3}}^{2}-y_{k_{3}}^{2}} & \frac{2\sqrt{S_{1}(S_{1}-l_{k_{1}k_{2}})(S_{1}-l_{k_{1}k_{3}})(S_{1}-l_{k_{2}k_{3}})}}{l_{k_{1}k_{2}}}
\end{bmatrix},\\
\mathbf{x}=&\begin{bmatrix}x_{O^{'}}\\
y_{O^{'}}
\end{bmatrix},\;
\mathbf{b}=\begin{bmatrix}\frac{1}{2}(r_{k_{1}}^{2}-r_{k_{2}}^{2}+l_{k_{1}k_{2}}^{2})\\
\frac{1}{2}(r_{k_{1}}^{2}-r_{k_{3}}^{2}+l_{k_{1}k_{3}}^{2})
\end{bmatrix}
\end{aligned}
\end{equation}

To verify the trilateration method improves the overall Kinect measurements, three
calculated positions of $O$ (i.e., $O_{k_{1}}(x_{O_{k_{1}}},\, y_{O_{k_{1}}}),\: O_{k_{2}}(x_{O_{k_{2}}},\, y_{O_{k_{2}}}),\: O_{k_{3}}(x_{O_{k_{3}}},\, y_{O_{k_{3}}})$) are considered
\begin{equation}
\label{eq:relocate}
\left\{
\begin{array}{l}
O_{k_{1}}:x_{O_{k_{1}}}=\frac{D_{k_{1}}}{tan\theta_{k_{1}}^{'}}, y_{O_{k_{1}}}=D_{k_{1}}\\
O_{k_{2}}:x_{O_{k_{2}}}=x_{k_{2}}-\frac{D_{k_{2}}}{tan\theta_{k_{2}}^{'}}, y_{O_{k_{2}}}=D_{k_{2}}\\
O_{k_{3}}:x_{O_{k_{3}}}=x_{k_{3}}-D_{k_{3}}\cdot tan\theta_{k_{1}}^{'}, y_{O_{k_{3}}}=y_{k_{3}}-D_{k_{3}}\\
\end{array}
\right.
\end{equation}
where $O_{k_{1}}$, $O_{k_{2}}$ and $O_{k_{3}}$ are calculated solely based on $k_{1}$, $k_{2}$ and $k_{3}$, respectively. For example, $O_{k_{1}}(x_{O_{k_{1}}},\, y_{O_{k_{1}}})$ is calculated with the Kinect $k_{1}$'s position $k_{1}(x_{k_{1}},\, y_{k_{1}})$ and its corresponding depth-angle pair $(\theta_{k_{1}}, \:D_{k_{1}})$ by geometry rules. Thus, totally four calculated positions of $O$ (i.e., $O_{k_{1}}$, $O_{k_{2}}$, $O_{k_{3}}$ and $O^{'}$) are compared.

\section{Results and Discussion}
\subsection{Accuracy Evaluation}
The following subsections explain the experiment results corresponding with the subsections in Section \ref{Kinect Accuracy Assessment}.

\subsubsection{Accuracy Distribution}
\begin{figure}[!t]
\begin{centering}
\includegraphics[width=9cm]{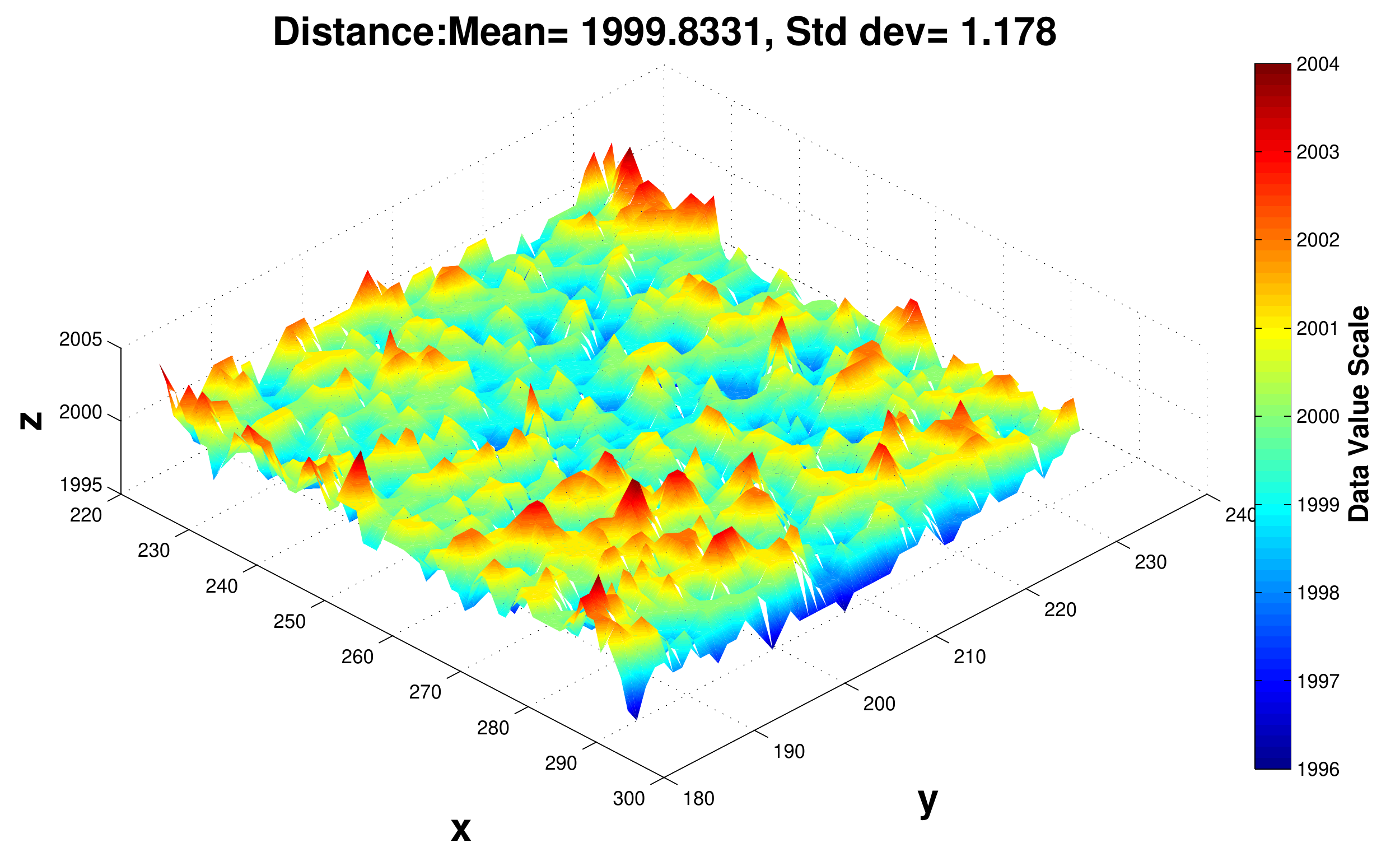}
\par\end{centering}
\protect\caption{\label{fig:Mean-depth-evaluation}Depth values' distribution representing the planar surface (70$\times$50 pixels)}
\end{figure}

\begin{figure*}[!t]
\begin{centering}
\includegraphics[width=17cm]{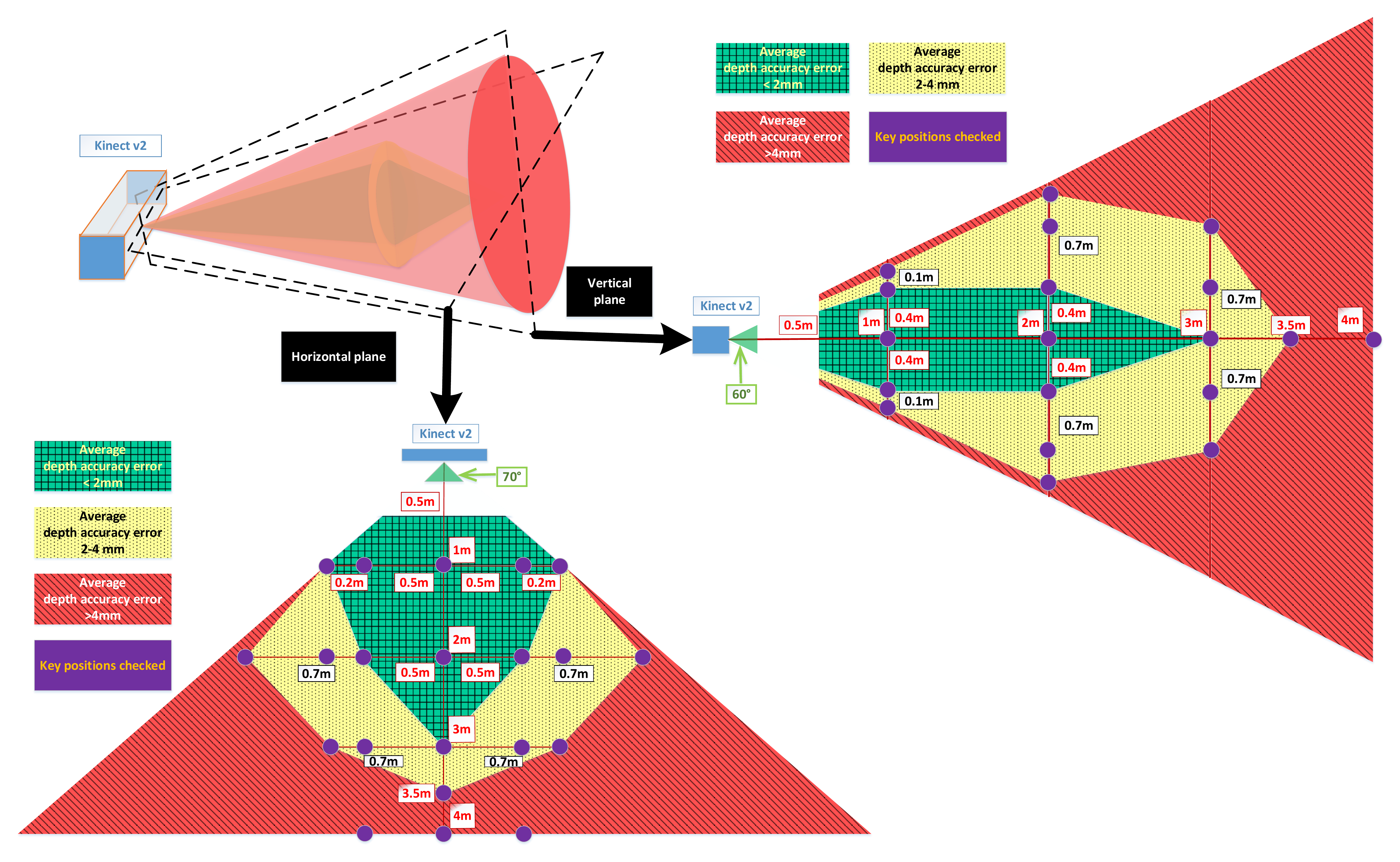}
\par\end{centering}

\protect\caption{\label{fig:Precision-Distribution}Accuracy error distribution of Kinect for Windows v2.}
\end{figure*}

Each frame is recorded when the planar surface is positioned at one of the key positions (\figurename\ref{fig:Precision-Distribution}). Totally there are 21 key positions in the horizontal plane and 19 positions in the vertical plane. Based on the depth values representing the screen in each recorded depth frame, the mean depth value, standard deviation and depth range are calculated and visualized. For example, \figurename\ref{fig:Mean-depth-evaluation} shows the planar surface when being pointed perpendicular towards Kinect v2 (the central key position at 2m). The x and y axis represent horizontal and vertical axis of the frame (the depth frame contains a 512$\times$424 pixel matrix where the x values range from 1 to 512 and y values range from 1 to 424) and z axis represents the depth value of the corresponding pixel. It is shown that the depth values range from 1996mm to 2004mm here. The mean depth value and its derivation are 1999.8mm and 1.2mm, respectively.

By summarizing the mentioned measurement accuracy of Kinect at all the key positions, we find that the accuracy error distribution of Kinect v2 satisfies an elliptical cone with 60 degrees' angle in vertical direction and 70 degrees' angle in horizontal direction (\figurename\ref{fig:Precision-Distribution}). Moreover, we use three colors (i.e., green, yellow and red) to indicate the different accuracy areas by dividing the space of accuracy error distribution into three regions. Specifically, in the green, yellow and red areas of both the horizontal and vertical plane, the average accuracy is less than 2mm, between 2mm and 4mm, and more than 4mm, respectively.

\subsubsection{Depth Resolution}
\begin{figure}
\begin{centering}
\subfloat[\label{fig:Resolution_Mean}]{\begin{centering}
\includegraphics[width=8cm]{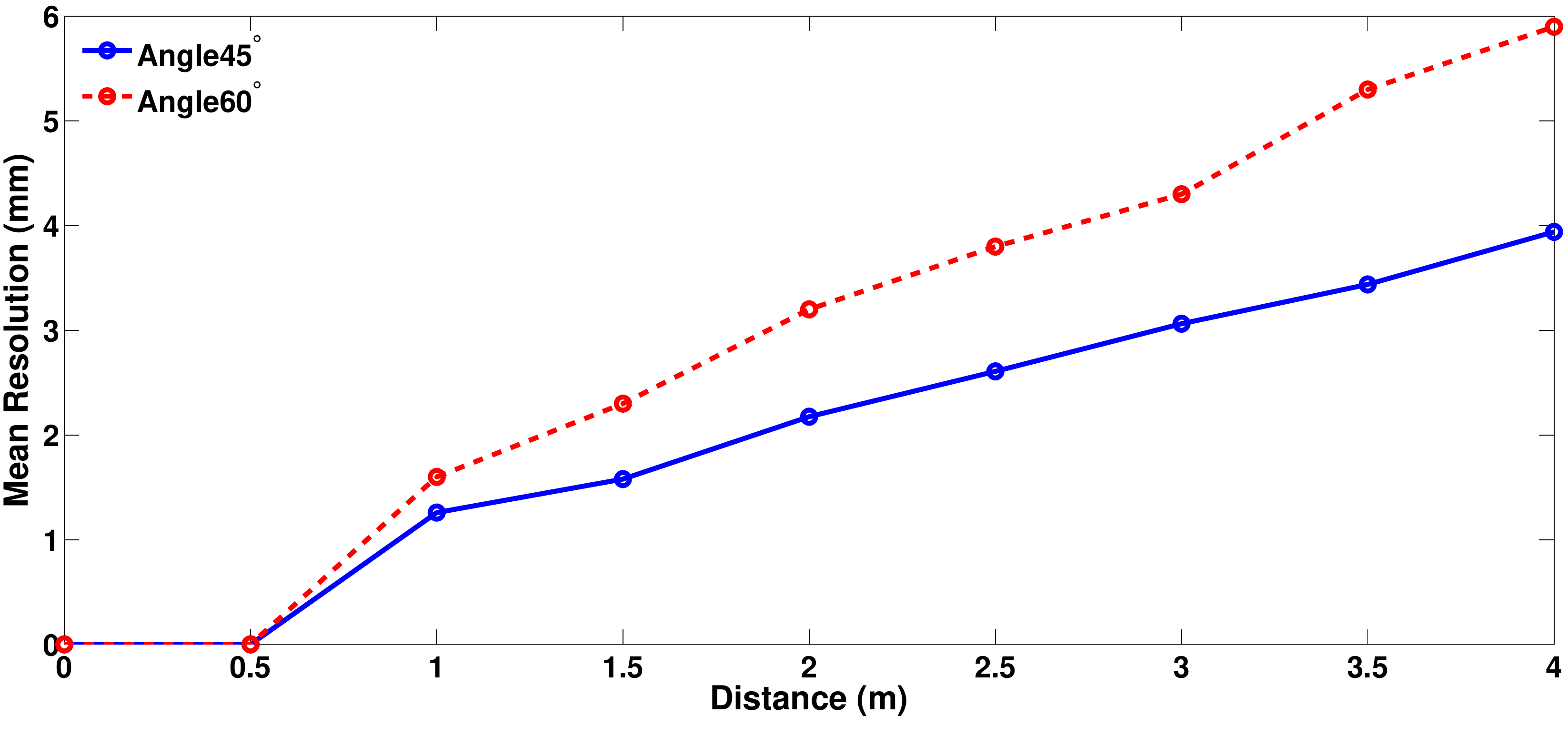}
\par\end{centering}
}
\par\end{centering}

\begin{raggedright}
\subfloat[\label{fig:Resolution_Stddev}]{\begin{centering}
\includegraphics[width=8.25cm]{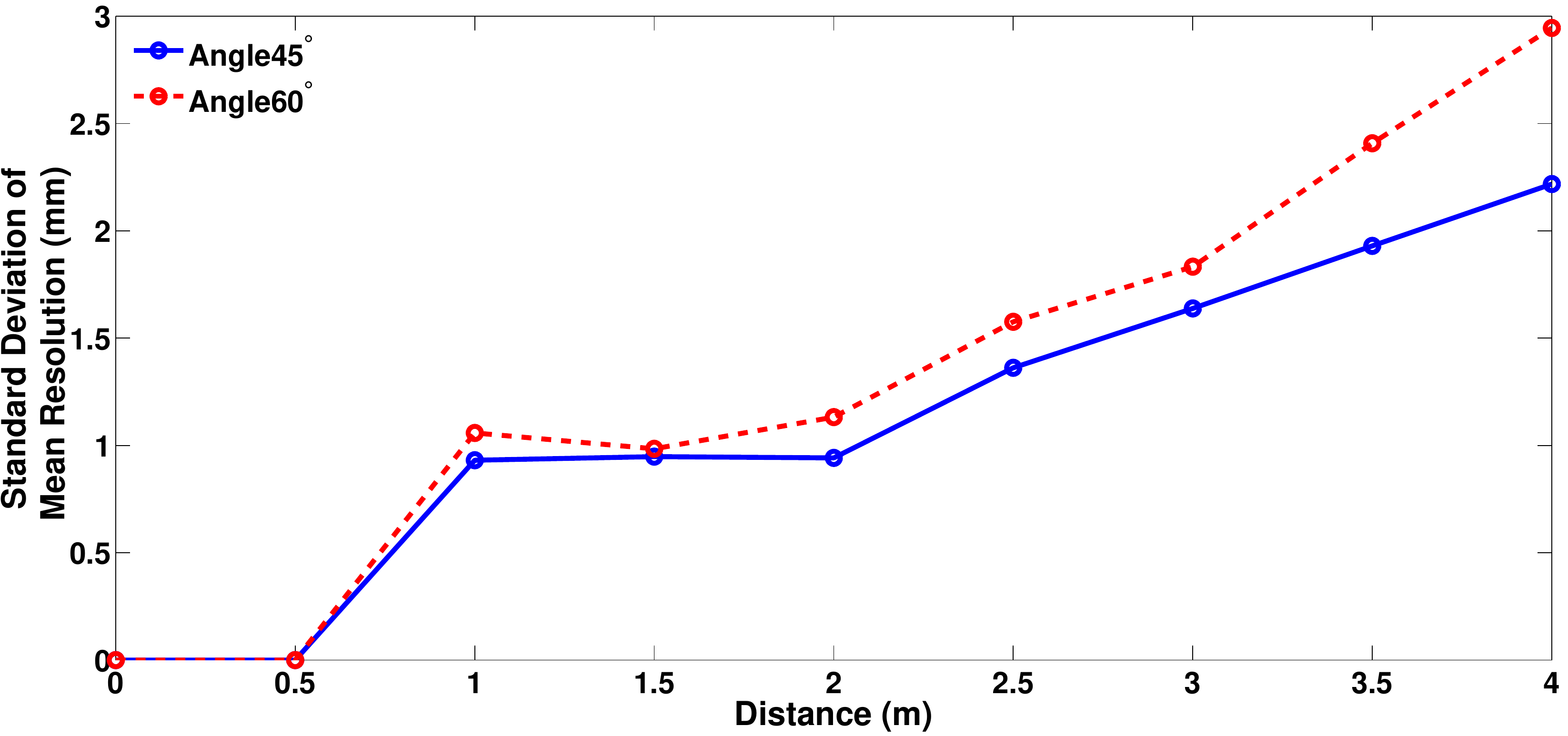}
\par\end{centering}
}
\par\end{raggedright}

\begin{centering}
\subfloat[\label{fig:Mean-depth-evaluation-3}]{\begin{centering}
\includegraphics[width=8cm]{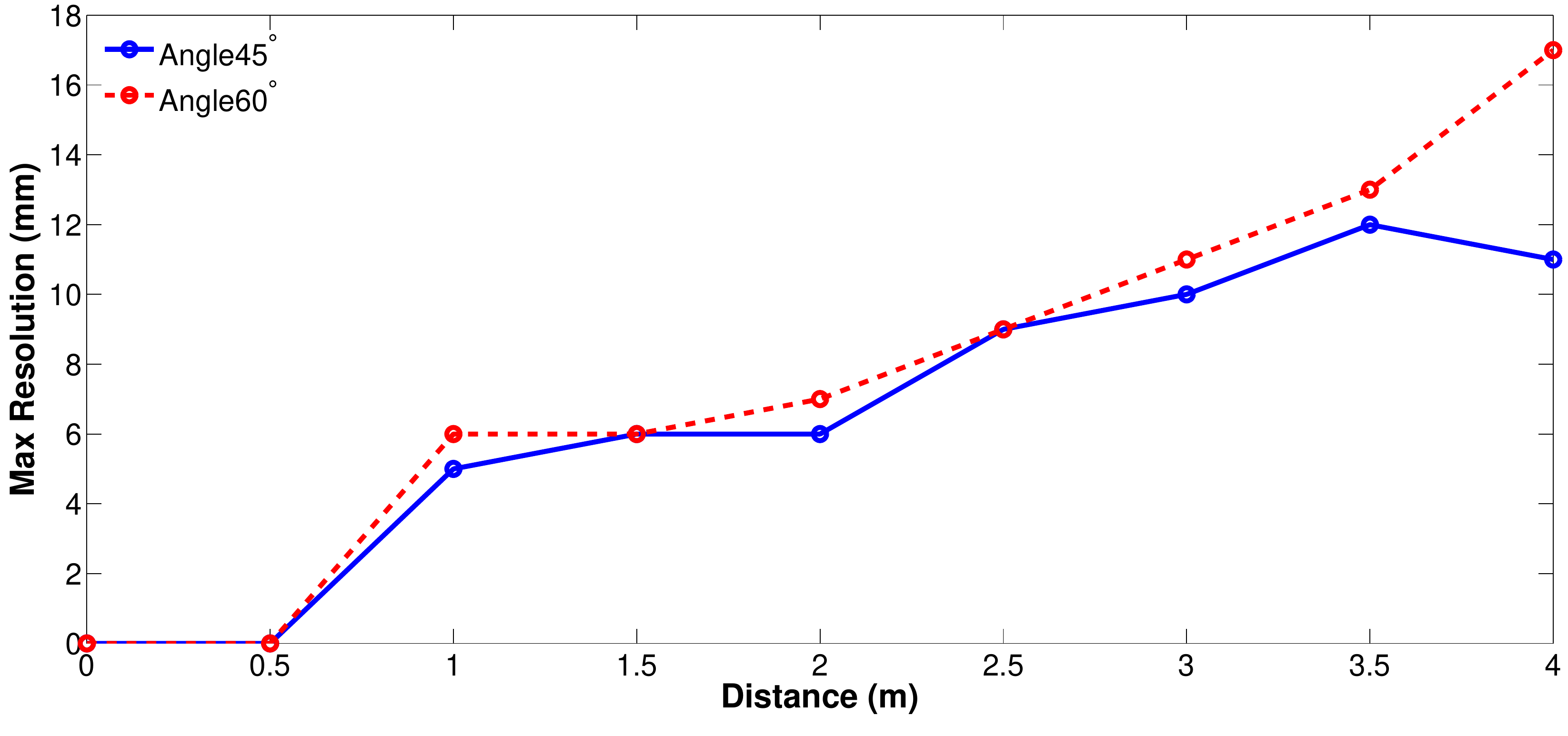}
\par\end{centering}
}\protect\caption{\label{fig:Resolution-Tendency}Resolution tendency with the Kinect's location and attitude. (a) Mean resolution tendency. (b) Standard deviation of the resolution tendency. (c) Max resolution tendency.}
\par\end{centering}
\end{figure}

As Section \ref{Kinect Accuracy Assessment} mentioned, the planar surface is pointed with two specific angles (45 degrees and 60 degrees) towards Kinect v2 at different distances. For each recorded depth frame, a depth difference between every two adjacent pixels is calculated and the mean resolution is computed by averaging all the mentioned depth difference between each adjacent pixels in the whole depth image. Similarly, the standard deviation and max resolution are also calculated. The tendency of mean resolution, max resolution and standard deviation are shown in \figurename\ref{fig:Resolution-Tendency} when the distance is increasing, where the horizontal axis represents the distance between the planar surface and the Kinect v2. Here, the solid line and dash line correspond with the tendency when the planar surface is inclined 45 and 60 degrees from Kinect v2, respectively. As Kinect v2 is not able to measure the depth within 0.5m, the resolution within 0.5m is set as 0 here. It is shown that: 1) the mean depth resolution and max resolution increase with distance, meaning that a coarser depth image is made with distance increase;  2) A larger tilt angle leads to a lower depth resolution and a larger standard deviation; 3) When the distance increases larger than 2m, the max resolution and standard deviation (for both 45 and 60 degrees tilt) increase faster.

\subsubsection{Depth Entropy}
To illustrate the phenomenon that the depth value of each pixel keeps varying within 6mm over time, at each position of the planar plane (1m, 2m, 3m or 4m away from the Kinect v2), totally 30 depth frames are recorded. For these recorded 30 frames, the sizes of each frame's pixel matrix representing the planar surface are the same, but the distribution of the pixels' values is different (depth value keep varying over time). For each two adjacent frame of the 30 frames, every variance between the two pixels at the same position of the two pixel matrixes is calculated, which means totally 29 variances are calculated for one specific position of the 30 pixel matrixes. Based on the calculated variance, an entropy distribution figure can be built by Equation \eqref{eq:eq1}. We found there is no apparent singularity in these entropy distribution figures. For example, \figurename\ref{fig:Entropy,-Entropy-Stdev,}(a) shows the entropy distribution when the planar plane is 2m away from the Kinect v2 where each pixel's value represents a specific corresponding pixel's entropy on the depth frames. Furthermore, the general tendency of mean entropy tendency and its standard deviation are analyzed in \figurename\ref{fig:Entropy,-Entropy-Stdev,} (b-c)) where the horizontal axis represents the distance between the Kinect v2 and the planar plane. It is shown that 1) the standard deviation of the entropy decreases when the distance increases because the number of the pixels representing the screen decreases with the distance increases; 2) the mean entropy increases fast after the distance is beyond 2m.

\begin{figure}
\begin{centering}
\subfloat[\label{fig:Depth-precision-at}]{\begin{centering}
\includegraphics[width=8cm]{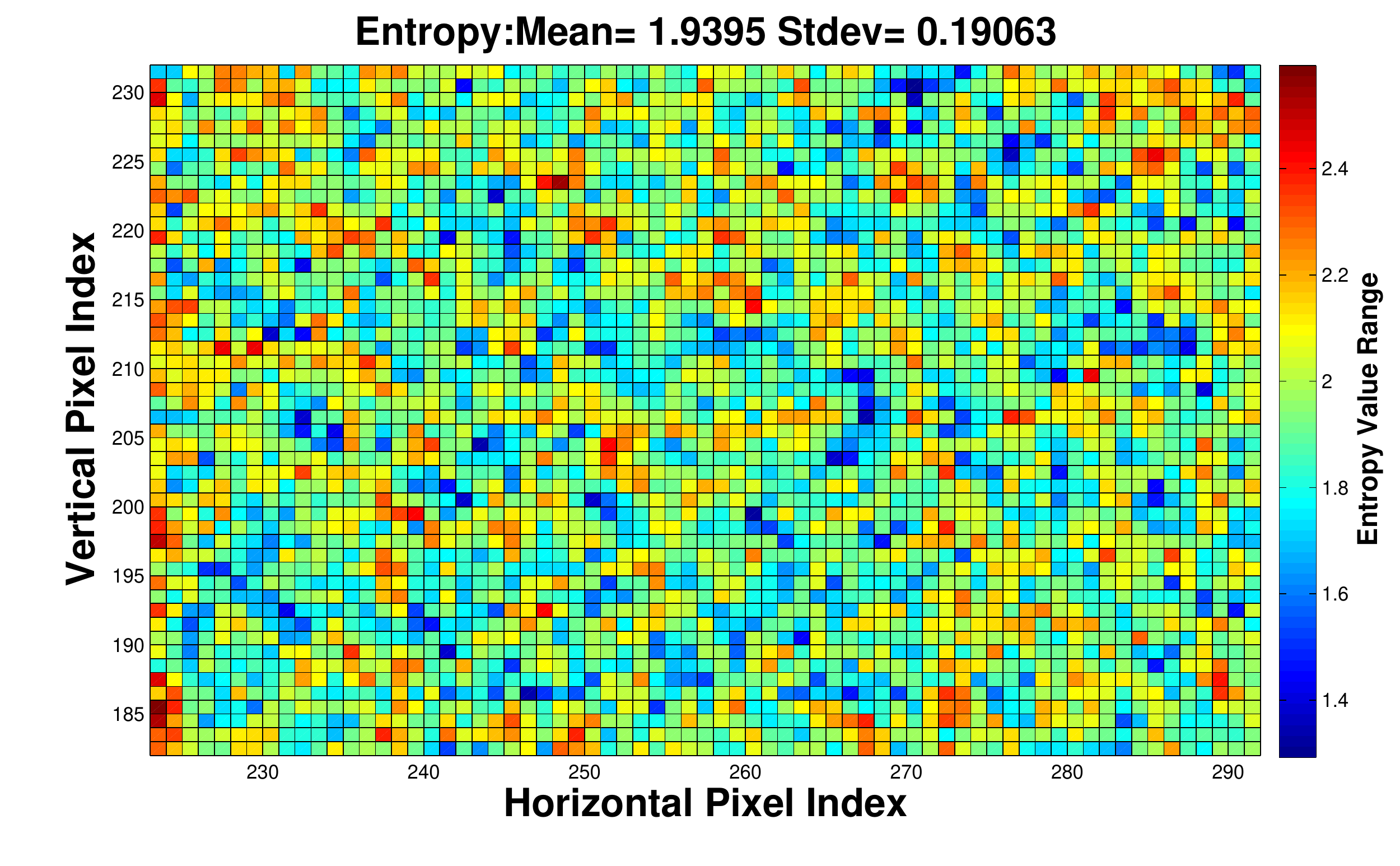}
\par\end{centering}
}
\par\end{centering}
\begin{centering}
\subfloat[\label{fig:EntropyTendency_Mean}]{\begin{centering}
\includegraphics[width=8cm]{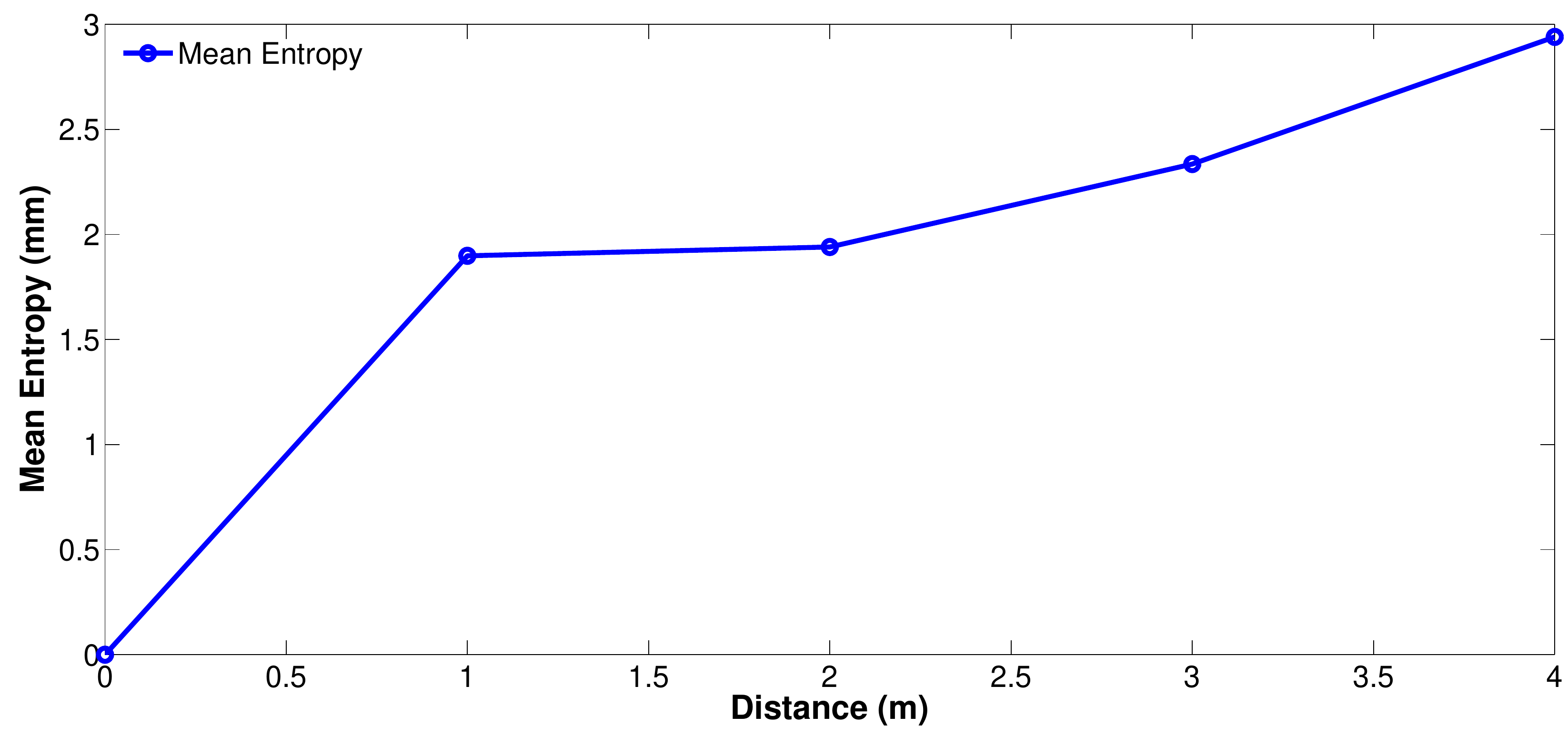}
\par\end{centering}
}
\par\end{centering}
\begin{centering}
\subfloat[\label{fig:EntropyTendency_stdev}]{\begin{centering}
\includegraphics[width=8cm]{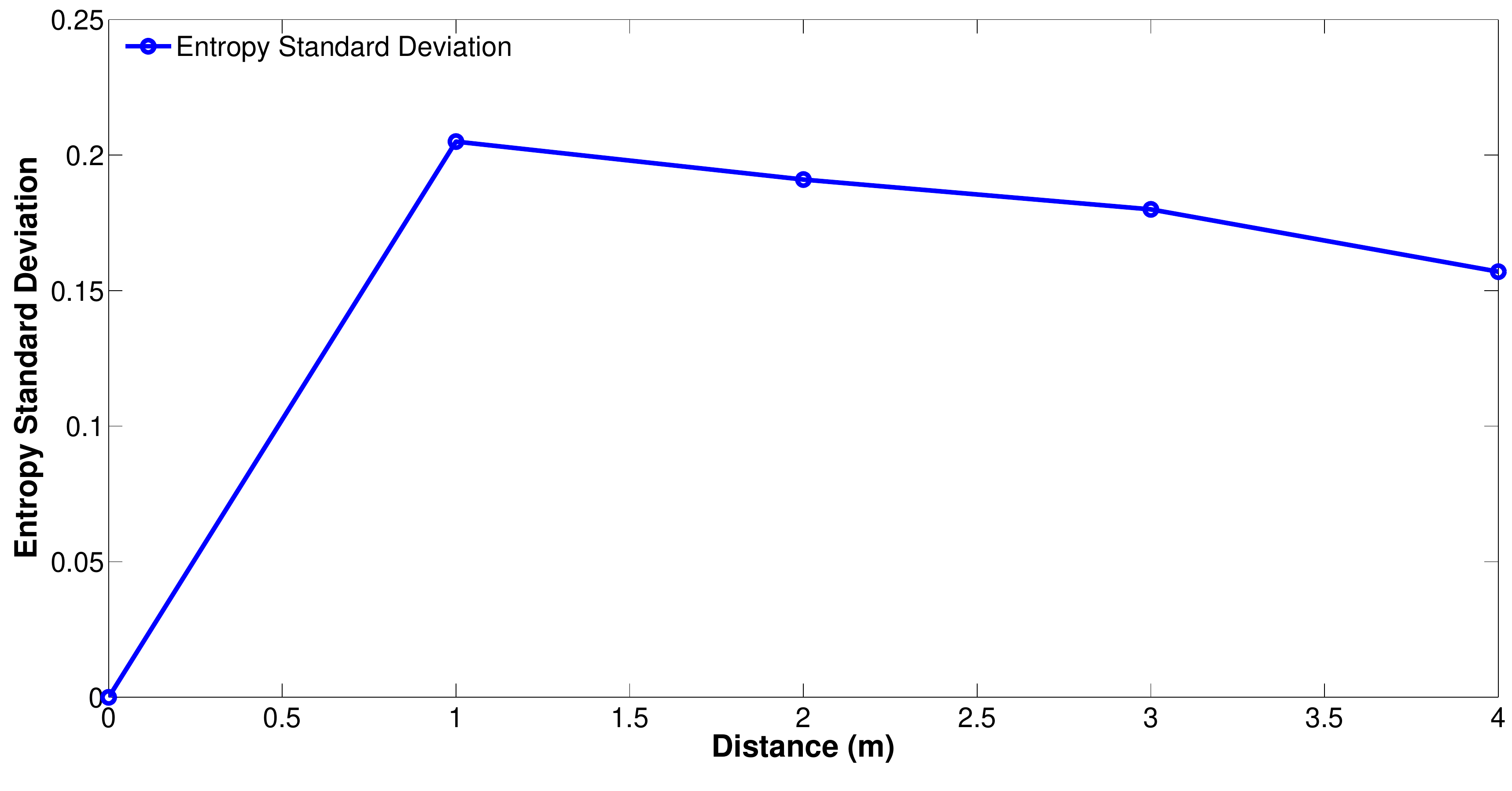}
\par\end{centering}
}\protect\caption{\label{fig:Entropy,-Entropy-Stdev,}Depth entropy of depth pixels. (a) Entropy distribution. (b) Mean entropy. (c) Standard deviation of the mean entropy.}
\par\end{centering}
\end{figure}

\subsubsection{Edge Noise}
\begin{figure}[!t]
\begin{centering}
\includegraphics[width=8cm]{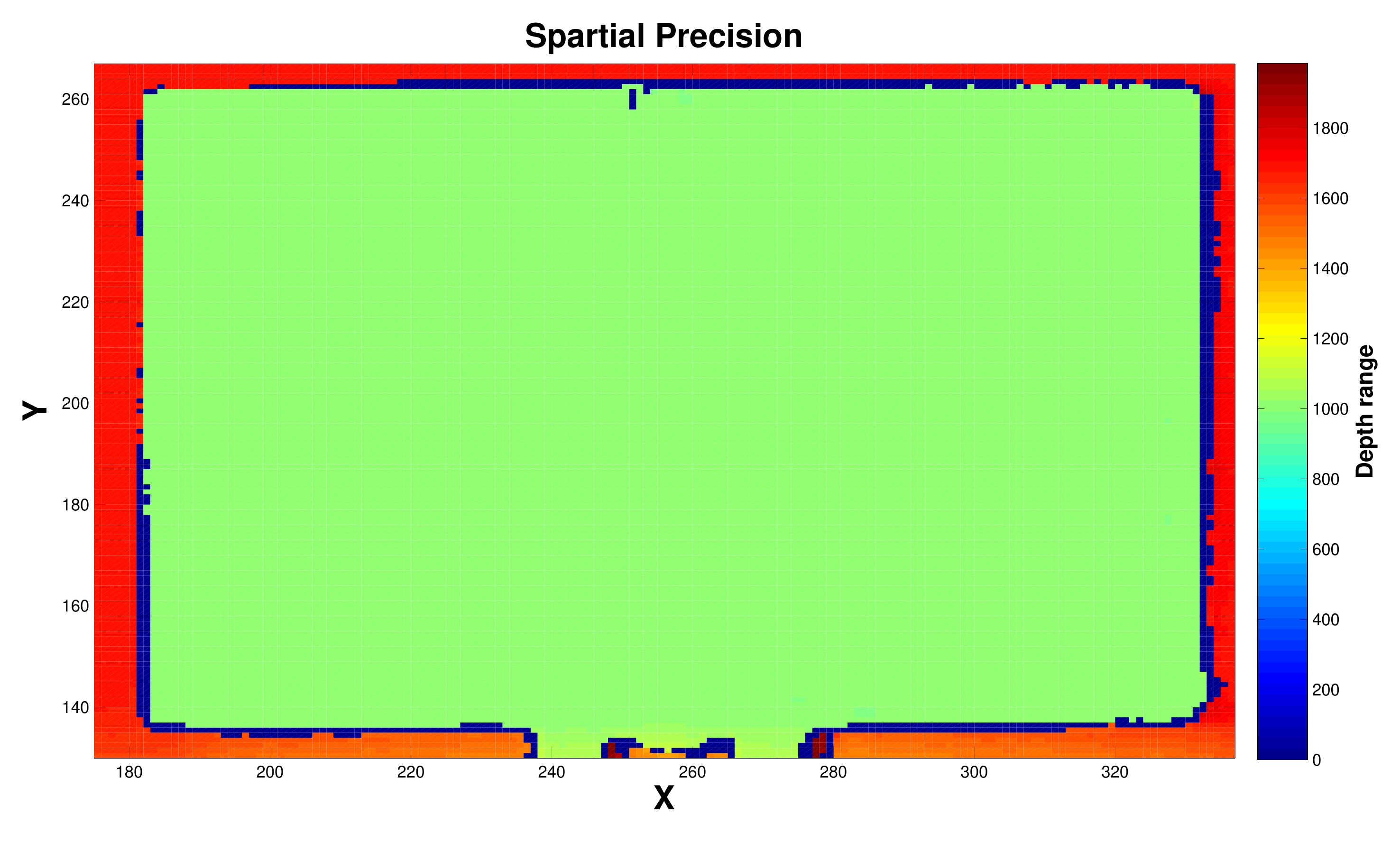}
\par\end{centering}
\begin{centering}
\protect\caption{\label{fig:Spartial-Precision-at}Zero-value contour of the measured planar plane when being set 1m from the Kinect v2.}
\par\end{centering}
\end{figure}

In real applications, wrongly measured pixels by Kinect v2 can lead to abnormal contours. Here, we use this phenomenon to evaluate the edge noise of Kinect v2. Specifically, the planar plane is set in front of the Kinect v2 and contours of the plane is analyzed. \figurename\ref{fig:Spartial-Precision-at} shows a depth frame when the planar plane is located 1m away from the Kinect v2 where x and y axis represent the horizontal and vertical index of the frame, respectively.

We observe that zero-value contours existing around the planar plane's edges (shown as deep blue contour). According to the manual of Kinect v2, a zero depth value means the position observed is out of Kinect sensor's observation range, but apparently the edges of this planar is within the measurable range (0.5m-4m). Here, the width of the zero pixels is 1\textasciitilde{}2 pixels (roughly 2-4mm) when the planar surface is 1m away from Kinect v2. According to our experiments, all the objects, with at least 0.5m distance from its background, have zero contours with a maximum width of 2 pixels. It is noted that there is no apparent increase of the width of the zero value contour when the distance from the Kinect v2 to object increases (from 0.5m to 4m).

\subsubsection{Structural Noise}

\begin{figure}[!t]
\begin{centering}
\includegraphics[width=8cm]{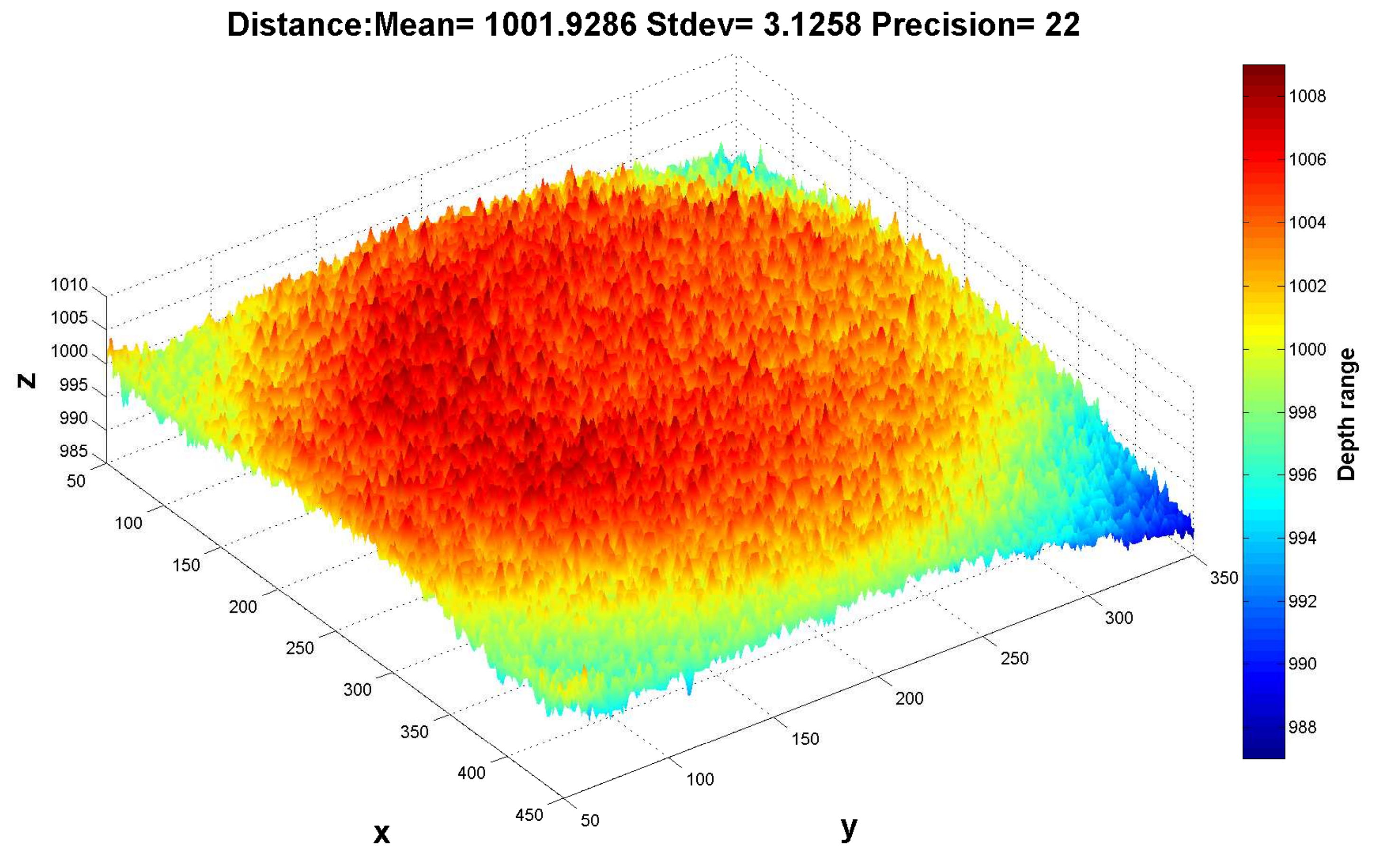}
\par\end{centering}

\protect\caption{\label{fig:Depth-distribution-1}Ring shape of the captured planar plane when being set 1m from the Kinect v2 (450$\times$400 pixels).}

\end{figure}

Kinect v2 has structural noise when capturing depth images. To analyze this issue, we test the measured flat plane in front of Kinect at 1m distance as shown in \figurename\ref{fig:Depth-distribution-1} where x and y axis represent the horizontal and vertical index of the frame respectively. It can be observed that the recorded depth values are distributed in the shape of rings, i.e., the depth values of pixels decrease when their distance to the central part increases. The reason of this ring phenomenon is hard to understand. However, we think it might be due to the diffraction coming from the random variance of the depth pixels mentioned in Subsection \ref{Depth Entropy}. 

\subsection{Multi-Kinect Trilateration Results}

As the multi-Kinect trilateration setup described in Section IV, three Kinect v2 are positioned at each vertex of a isosceles triangle. Without loss of generality, we set the location of $O$ in front of Kinect v2 $k_{3}$. The radiuses ($r_{1}$, $r_{2}$ and $r_{3}$) and the vertexes positions ($k_{1}(x_{k_{1}},\:y_{k_{1}})$, $k_{2}(x_{k_{2}},\:y_{k_{2}})$, $k_{3}(x_{k_{3}},\:y_{k_{3}})$) are calculated first and then the optimized position by trilateration method($O^{'}$) is computed by the steps described in Section IV as shown in Table \ref{tab:Trilateration-variables}. 

\begin{table}
\protect\caption{\label{tab:Trilateration-variables} Localization Comparison between by Single Kinect and by Multi-Kinect Trilateration}
\begin{centering}
\begin{tabular}{|>{\centering}p{2cm}|>{\centering}p{2cm}|>{\centering}p{3.5cm}|}
\hline 
{\scriptsize{}Method} & {\scriptsize{}Measured/Calculated Position of $O$} & {\scriptsize{}Measurement/Calculation Error (Difference between  $O_{o}$ and the Measured/Calculated Position of $O$) }\tabularnewline
\hline 
\hline 
{\scriptsize{}Measurement from Kinect $k_{1}$}  & {\scriptsize{}$O_{k_{1}}$}{\scriptsize \par}

{\scriptsize{}(2197.1, 4016.4)} & {\scriptsize{}$|O_{0}-O_{k_{1}}|=64.23\:mm$}\tabularnewline
\hline 
{\scriptsize{}Measurement from Kinect $k_{2}$}  & {\scriptsize{}$O_{k_{2}}$}{\scriptsize \par}

{\scriptsize{}(2057.4, 3985.0)} & {\scriptsize{} $|O_{0}-O_{k_{2}}|=78.99\:mm$}\tabularnewline
\hline 
{\scriptsize{}Measurement from Kinect $k_{3}$}  & {\scriptsize{}$O_{k_{3}}$}{\scriptsize \par}

{\scriptsize{}(2135.0, 3989.4)} & {\scriptsize{}$|O_{0}-O_{k_{3}}|=10.66\:mm$}\tabularnewline
\hline 
{\scriptsize{}Multi-Kinect Trilateration} & {\scriptsize{}$O^{'}$}{\scriptsize \par}

{\scriptsize{}(2156.4, 4013.9)} & {\scriptsize{}$|O_{0}-O^{'}|=25.61\:mm$}\tabularnewline
\hline 
\end{tabular}

\begin{tablenotes}
            \item[*]* The measured position of $O_{o}$ is (2135, 4000).
\end{tablenotes}

\par\end{centering}
\end{table}

To verify the validity of the multi-Kinect trilateration, the position of $O$ (denoted as $O_o$) is measured. $|O_{k_{1}}-O_{o}|$, $|O_{k_{2}}-O_{o}|$ and $|O_{k_{3}}-O_{o}|$ are compared with $|O^{'}-O_{o}|$ where $O_{k_{1}}$, $O_{k_{2}}$ and $O_{k_{3}}$ are obtained solely based on the measurements from Kinect $k_{1}$ to Kinect $k_{3}$, respectively. $O^{'}$ is computed based on the measurements from the three Kinects. The four mentioned measurement errors are compared, which shows that the position of $O$ observed by Kinect $k_{3}$ has smallest measurement error (10.66 mm); the position of $O$ observed by Kinect $k_{1}$ and $k_{2}$ has large measurement error (64.23 mm for Kinect $k_{1}$ and 78.99 mm for Kinect $k_{1}$); while the position of $O$ observed by the proposed multi-Kinect trilateration method has the measurement error in between, respectively.

The measurement of $O$ from Kinect  $k_{3}$ has only 10 mm error because it is set straightly towards the plannar surface while the other two Kinects are set with a 60-degree angle towards the planar plane. Although the measurement of Kinect $k_{3}$ is more accurate here, the observed object cannot be always at the central area of the Kinect sensor's view range. In this case, the trilateration method has an overall good performance. Besides, the measurement result computed by multi-Kinect trilateration is better than the mean measurement error (51.22 mm) of the three Kinect sensors.

\section{Conclusion}

Compared with Kinect v1, the newly released Kinect for Windows
sensor v2 has improved performances on the hardware according to the published coefficients on the official website. In this
paper, we investigated several properties that are important
for practical usage of Kinect v2, including accuracy distribution, depth resolution, depth entropy, edge noise and structural noise. Based on the results we obtained from the experiments, Kinect v2 has good accuracy if the object is positioned
within the green regions (\figurename\ref{fig:Precision-Distribution}). Furthermore, we proposed a muti-Kinect
trilateration approach to improve the accuracy when three Kinect sensors are used. The experiment shows that the multi-Kinect trilateration method works well even when the sensors are positioned more than four meters away from the target object.

There are also some other parameters affecting the Kinect v2's performance. For example,
an object with reflective material (like mirror) can lead to a problem that the
IR (Infrared) light emitted by Kinect sensor cannot be reflected back to the camera,
making the depth values unreliable or unable to be determined.
Similarly, an object covered with light-absorbing materials (like carbon black) can
cause less IR light reflected back to the camera. Furthermore,
if two Kinect v2 sensors are positioned towards each other, the area around
the camera would disappear in the depth images captured by both two
sensors. This phenomenon happens probably because the camera interference IR light from the other sensor. The
unattractive power bricks, complex cables and the high requirement for the laptop also limit the use of the sensor.

In summary, as a depth sensor with a relatively low price (much lower
than the professional depth cameras or tracking systems like Vicon), Kinect for
Windows sensor v2 shows acceptable performance which has a great potential to be applied to many fields, such as entertainment, education and medicine areas, and so on.

\ifCLASSOPTIONcaptionsoff
  \newpage
\fi

\section*{Acknowledgment}
The authors extend their appreciation to the Deanship of Scientific Research at King Saud University, Riyadh, Saudi Arabia for funding this work through the International Research Group Program (No. IRG14-30).



%

\bibliographystyle{unsrt}
\bibliography{Bibliov}


%

\begin{IEEEbiography}[{\includegraphics[width=1in,height=1.25in,clip,keepaspectratio]{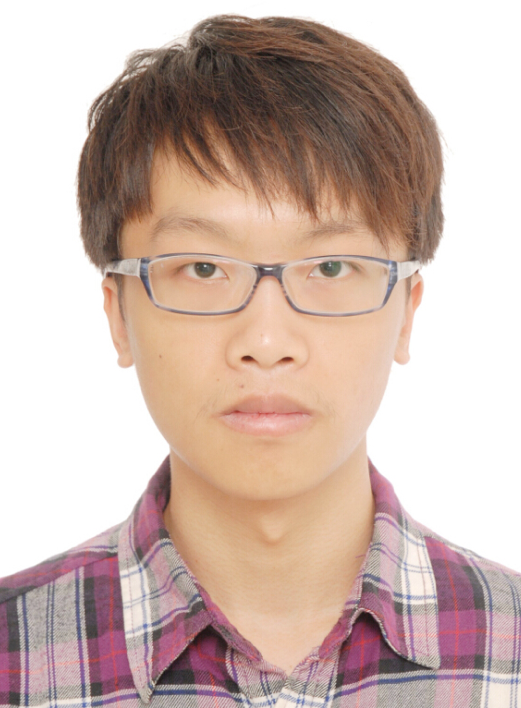}}]{Lin Yang}
received his B.Eng. in Software Engineering from Sichuan University in 2013 and started his M.Sc in University of Ottawa at the same year. He is currently working on the project, Touchable Avatar, at MCR lab of University of Ottawa, creating a touchable and interactive hologram to bring benefit to remote communication. His another research topic is 3D Sensing and Tracking of Human Gait Movement, finding a solution to improve the accuracy of Microsoft Kinect for Windows sensor v2 for medical purpose. His research interests include computer graphics, machine learning and human-machine interaction.
\end{IEEEbiography}
\vfill
\newpage
\begin{IEEEbiography}[{\includegraphics[width=1in,height=1.25in,clip,keepaspectratio]{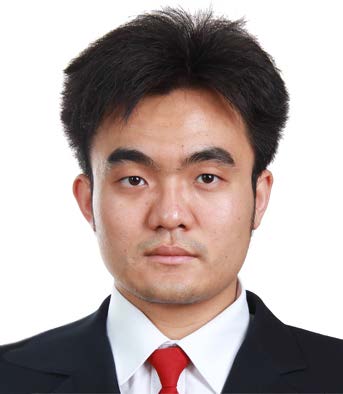}}]{Longyu Zhang}
received his M.A.Sc degree in Electrical and Computer Engineering from University of Ottawa, and B.Eng degree in Telecommunications Engineering from Tianjin Polytechnic University in 2012 and 2010 respectively. He is currently pursuing his Ph.D degree in Electrical and Computer Engineering at Multimedia Communications Research Laboratory (MCRLab), University of Ottawa. His research interests focus on haptics and computer vision.  He is a student member of IEEE.
\end{IEEEbiography}


\begin{IEEEbiography}[{\includegraphics[width=1in,height=1.25in,clip,keepaspectratio]{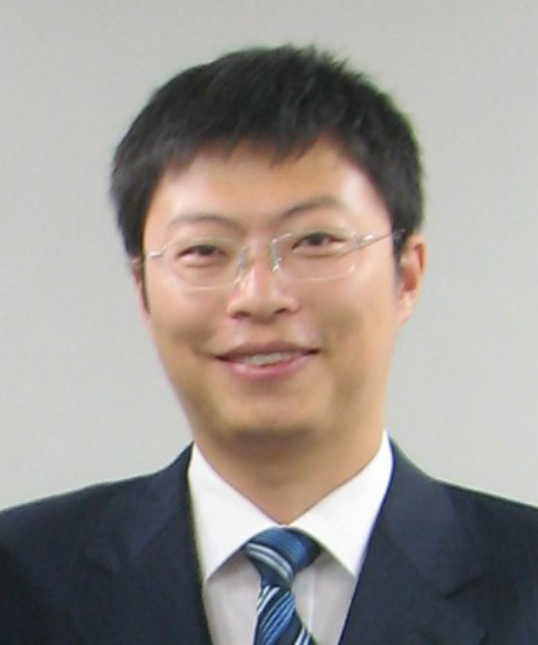}}]{Haiwei Dong}
received Dr.Eng. in Computer Science and Systems Engineering and M.Eng. in Control Theory and Control Engineering from Kobe University (Japan) and Shanghai Jiao Tong University (P.R.China) in 2010 and 2008, respectively. He is currently with University of Ottawa. Before that, he was appointed as Postdoctoral Fellow in New York University AD; Visiting Scholar in the University of Toronto; Research Fellow (PD) in Japan Society for the Promotion of Science (JSPS); Science Technology Researcher in Kobe University; Science Promotion Researcher in Kobe Biotechnology Research and Human Resource Development Center. His research interests include robotics, haptics, control and multimedia. He is a member of IEEE and ACM.
\end{IEEEbiography}


\begin{IEEEbiography}[{\includegraphics[width=1in,height=1.25in,clip,keepaspectratio]{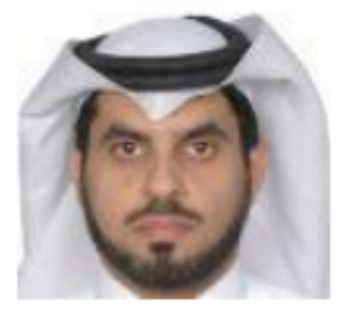}}]{Abdulhameed Alelaiwi}
is a vice dean for technical affairs, Scientific Research Deanship, King Saud University and a faculty member in Software Engineering Department, College of Computer and Information Sciences, KSU. He holds a Ph.D. in the field of Software Engineering from the department of software engineering, Florida Tech University, USA, 2002. He obtained his M.Sc. in the field of software engineering from Florida Tech University in 1998. He worked in the industry around 7 years, before joining King Saud University. He continues to consult with local corporations in the areas of Software Engineering, E-Government, and Information security. Dr. Alelaiwi has been consulting with many companies as well as government sites such as Ministry of Labor, Ministry of Defense, and Ministry of Information. His industry experience includes applications of the state-of-the-art techniques in solving software engineering, project engineering problems at Ministry of Labor, where he was Minister Consultant and the director of computer center at the same time. Before that, at Vinnell Corporation, Dr. Alelaiwi has developed complex algorithms and simulation programs in solving military related problems.
\end{IEEEbiography}

\begin{IEEEbiography}[{\includegraphics[width=1in,height=1.25in,clip,keepaspectratio]{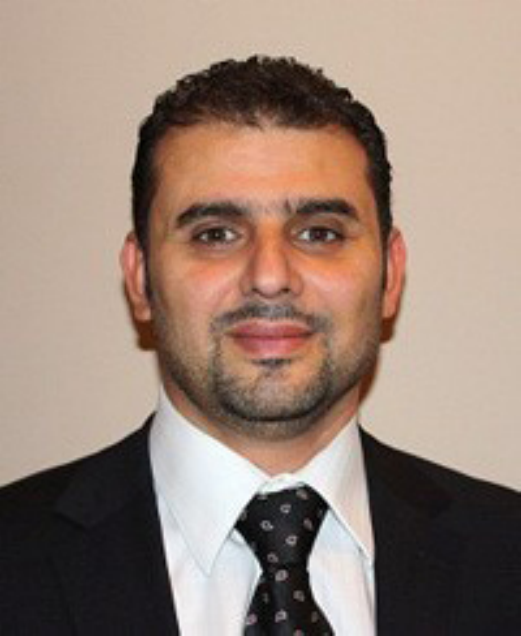}}]{Abdulmotaleb El Saddik}
is University Research Chair and Distinguished Professor in the School of Electrical Engineering and Computer Science at the University of Ottawa. He held regular and visiting positions in Canada, Spain, Saudi Arabia, UAE, Germany and China. He is an internationally-recognized scholar who has made strong contributions to the knowledge and understanding of multimedia computing, communications and applications. He has authored and co-authored four books and more than 400 publications, chaired more than 40 conferences and workshops and has received research grants and contracts totaling more than \$18M. He has supervised more than 100 researchers. He received several international awards among others ACM Distinguished Scientist, Fellow of the Engineering Institute of Canada, and Fellow of the Canadian Academy of Engineers and Fellow of IEEE and IEEE Canada Computer Medal.
\end{IEEEbiography}
\vfill


\end{document}